\documentclass[floats,longbibliography,twocolumn,prl,reprint]{revtex4-1}
\usepackage{graphicx}
\usepackage{bm}
\usepackage{color}
\usepackage{mathtools}

\newcommand{\bk}{{\bf k}}
\newcommand{\bq}{{\bf q}}

\newcommand{\sigmab}{\overline\sigma}

\begin{document}
\title{Possible Flexoelectric Origin of the Lifshitz Transition in LaAlO$_3$/SrTiO$_3$ Interfaces}
\author{Amany Raslan}
\author{W. A. Atkinson}\email{billatkinson@trentu.ca}
\affiliation{Department of Physics \& Astronomy, Trent University, Peterborough Ontario, Canada, K9L 0G2}
\date{\today}
\begin{abstract}
Multiple experiments have observed a sharp transition in the band structure of LaAlO$_3$/SrTiO$_3$ (001) interfaces as a function of applied gate voltage.  This Lifshitz transition, between a single occupied band at low electron density and multiple occupied bands at high density, is remarkable for its abruptness.  In this work, we propose a mechanism by which such a transition might happen.  We show via numerical modeling that the simultaneous coupling of the dielectric polarization to the interfacial strain (``electrostrictive coupling'') and strain gradient (``flexoelectric coupling'') generates a thin polarized layer whose direction reverses at a critical density.  The Lifshitz transition occurs concomitantly with the polarization reversal and    is first-order at $T=0$.  A secondary Lifshitz transition, in which electrons spread out into semiclassical tails, occurs at a higher density.  
\end{abstract}
\maketitle


LaAlO$_3$ (LAO) and SrTiO$_3$ (STO) are band insulators; however, when four or more monolayers of LAO are grown on top of a STO substrate, a mobile two-dimensional electron gas (2DEG) forms on the STO side of the interface \cite{Ohtomo:2004hm,Thiel:2006eo}.  One compelling feature of these interfaces is that  the character of the 2DEG  changes dramatically with the application of a gate voltage.    Indeed, for (001) interfaces there is a narrow doping range over which the superconducting transition temperature \cite{Caviglia:2008uh,Dikin:2011gl,Biscaras:2012vd,Maniv:2015cc,Hurand:2015cf}, the spin-orbit coupling \cite{Caviglia:2010jv,BenShalom:2010kv,Liang:2015fy,Hurand:2015cf}, and the metamagnetic response \cite{Joshua:2013wl} change by an order of magnitude.  Furthermore, the superconducting gap and resistive transition appear at different temperatures at low electron densities, $n_\mathrm{2D}$, but track each other closely at high $n_\mathrm{2D}$ \cite{Richter:2013gn}.  This qualitative distinction between low and high doping has also been seen in quantum dot transport experiments, which reveal a crossover from attractive to repulsive pairing interactions with increasing $n_\mathrm{2D}$ \cite{Cheng:2016tk}. While there is general agreement that the sensitivity to doping  is connected to an observed Lifshitz transition \cite{Biscaras:2012vd,Joshua:2012bl,Smink:2017cc,Niu:2017ih,Smink:2018tu} between a single occupied band at low density and multiple occupied bands at high density \cite{Kim:2013vz,Zhong:2013fv,Khalsa:2013hk,Fischer:2013gg,Zhou:2015uo,Nakamura:2013cb,Smink:2018tu,Maniv:2015cc,Nandy:2016jm},  the mechanism by which this transition happens is not  established. 
 
Density functional theory (DFT), while instrumental in establishing fundamental interface properties \cite{Popovic:2008ft,Son:2009wb,Pentcheva:2009ef,Stengel:2011hy}, finds electron densities that are an order of magnitude larger than the Lifshitz transition density $n_L\sim 0.02$--0.05 electrons per 2D unit cell,  and cannot easily be tuned through the  transition. Schr\"odinger-Poisson calculations, for which $n_\mathrm{2D}$ can be continuously tuned,  persistently find multiple occupied bands even for $n_\mathrm{2D} \ll n_L$ \cite{Biscaras:2012vd,Smink:2017cc,Khalsa:2012fu,Raslan:2017gh,Li:2018hy}.  Indeed, we showed previously that, because of STO is close to a ferroelectric (FE) quantum critical point\cite{Rowley:2014bda}, electrons become deconfined from an ideal interface as $n_\mathrm{2D}\rightarrow 0$, and form a dilute quasi-three-dimensional (quasi-3D) gas  extending far into the STO substrate \cite{Atkinson:2017jt}.  This result is incompatible with experiments and raises the question, why is  only a single band  occupied at low $n_\mathrm{2D}$?

Furthermore, the evolution of the band structure with $n_\mathrm{2D}$ is highly unusual.  Early work \cite{Joshua:2012bl} showed that the filling of the lowest band is constant for $n_\mathrm{2D}>n_L$, and recent experiments have found  geometries in which  its filling  decreases \cite{Smink:2017cc,Niu:2017ih,Smink:2018tu}.    In the latter case, the first band actually empties itself into higher energy bands with increasing chemical potential.  This cannot be understood within rigid-band models.

Moving beyond rigid bands, intra-atomic Coulomb (Hubbard) interactions \cite{Maniv:2015cc,Nandy:2016jm} have been invoked as a possible explanation for the unusual band filling \cite{Joshua:2012bl,Niu:2017ih}; however,  as we show below, these are too weak to be relevant. 
We suggest that a purely electronic explanation for the Lifshitz transition is unlikely, and focus instead on STO's unique dielectric properties.  In particular, STO's proximity to a FE transition allows for a large coupling  between the dielectric polarization and lattice strains \cite{Zubko:2013bt}.  This makes STO interfaces qualitatively different from conventional metallic interfaces, and as we show enables a novel switchable state involving the lattice polarization and the  2DEG.

Strains arising from direct application of pressure or from lattice misfits between a thin film and its substrate form the basis of ``strain engineering''.  By such methods, one can modify the dielectric response \cite{Uwe:1976fj,Wang:2000vy}, and even stabilize ferroelectricity \cite{Uwe:1976fj,Haeni:2004gj}, in STO.  Furthermore, strains applied to metallic LAO/STO interfaces can be used to tune their carrier density and subband occupations \cite{Bark:2011fo,Behtash:2016dt,Seiler:2018eo}.  We distinguish between these extrinsic bulk strains, and the intrinsic  strain that arises from $c$-axis relaxation (perpendicular to the interface) within a few nm of the LAO/STO interface.  Because the LAO cap layer is grown on top of the STO substrate, the transverse (parallel to the interface) strain in the STO vanishes; however, there is a longitudinal strain $\eta(z)  \sim 0.01$--0.03 extending four or five layers into the STO \cite{Willmott:2007ip,Lee:2016dj}. This strain couples to the polarization $P$ through an electrostrictive contribution \cite{Uwe:1976fj}, $-g_{11} \eta P^2$, and a flexoelectric contribution \cite{Zubko:2013bt}, $-f_{11}P\partial_z \eta$, to the lattice free energy.  

While flexoelectric effects are typically negligible in bulk materials, they appear to be a generic feature of nanometer structures in FEs---including domain walls \cite{Yudin:2013,Gu:2014iq},  grain boundaries \cite{Gao:2018kk}, and cracks \cite{Abdollahi:2015bn}---where the strain gradient $\partial_z\eta$ is often enormous.  Flexoelectricity  substantially affects the performance of nanocapacitors \cite{Majdoub:2009il}, and numerous proposals for flexo-mechanical devices have been made \cite{Zubko:2013bt}.  Whereas nearly all previous work has focused on insulating FEs, we show here that flexoelectricity fundamentally alters the 2DEG at STO interfaces.

\begin{figure}[tb]
\includegraphics[width=\columnwidth]{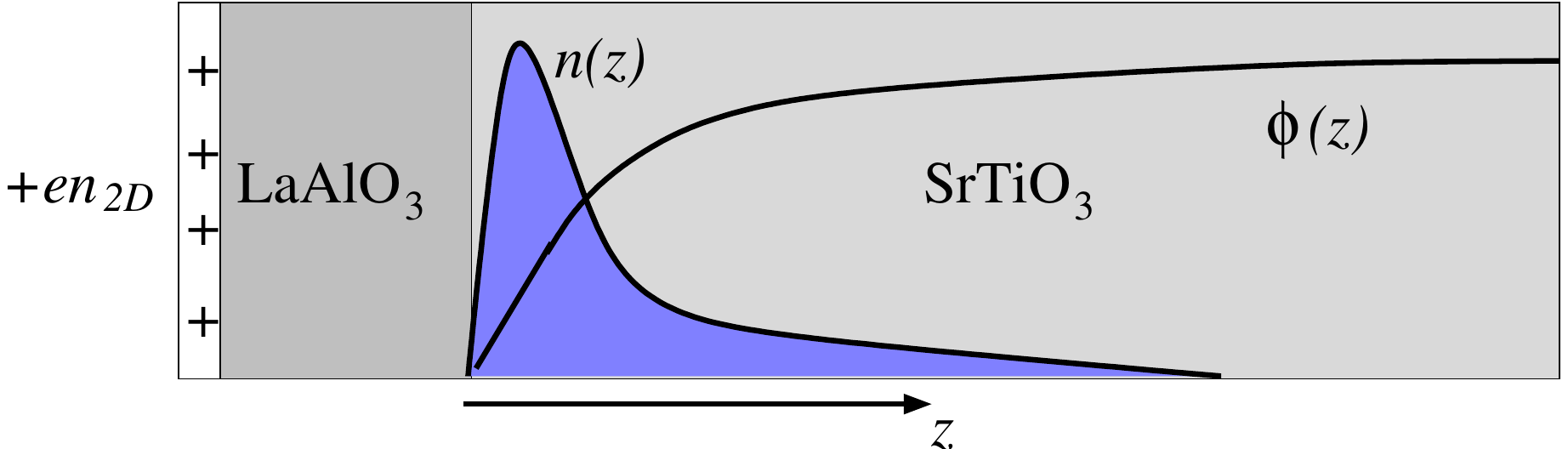}
\caption{Schematic LAO/STO interface showing the interface geometry, the electron potential energy  $\phi(z)$ and electron density $n(z)$.}
\label{fig:model}
\end{figure}

We consider a model (001) interface between a STO substrate and a  LAO cap layer [Fig.~\ref{fig:model}].  
Several doping mechanisms, including electronic reconstruction \cite{Nakagawa:2006gt}, a nonstoichiometric LAO surface\cite{Bristowe:2014fc,Piyanzina:2018ua}, and top-gating,  contribute to the 2DEG in the STO substrate.  For our purposes, these doping mechanisms can be modeled by a positive  charge density $en_\mathrm{2D}$ on the LAO surface, and  overall neutrality requires a 2D charge density of $-en_\mathrm{2D}$ in the STO substrate.  
The 2DEG forms on the STO side of the interface owing to the wide LAO band gap \cite{Gariglio:2015jx}, and is confined  by a potential $\phi(z)$ that depends on $n_\mathrm{2D}$ and on the dielectric screening in the substrate.  Doped electrons reside  on the Ti $t_{2g}$ orbitals.  

The $t_{2g}$ bands are obtained from a tight-binding Hamiltonian that has been fitted to Shubnikov-de Haas measurements in bulk STO \cite{Allen:2013wk}, and the  lattice polarization is obtained from a Landau-Devonshire free energy that has been fitted to bulk measurements of the dielectric function \cite{Raslan:2017gh} (see the Supplemental Material \cite{SM}). The polarization and electronic eigenstates are coupled through the electric field, which is obtained from the Poisson equation.   The model has a planar geometry, so that the polarization, electric field, etc., depend only on the distance $z$ from the interface; we discretize the model, so that there are $L$ layers of STO with, e.g.\ $P_i$ denoting the polarization in the $i$th layer.  

The Landau-Devonshire free energy is
\begin{equation}
{\cal U}=\frac{1}{2}\sum_{i,j} P_i {\tilde D}_{ij}P_j -  \sum_{i} \tilde E_i P_i + \mbox{quartic terms},
\label{eq:D}
\end{equation}
where $i$ and $j$ are layer indices, and $\tilde E_i$ and $\tilde D_{ij}$ are linear and quadratic coefficients of the free energy expansion.  The quartic terms in Eq.~(\ref{eq:D}), which are parameterized by a coefficient $\gamma$ \cite{SM}, are generally negligible in the doping range explored here \cite{Khalsa:2012fu,Raslan:2017gh}, but can be important when flexoelectric effects are included.  

\begin{figure}[tb]
\includegraphics[width=\columnwidth]{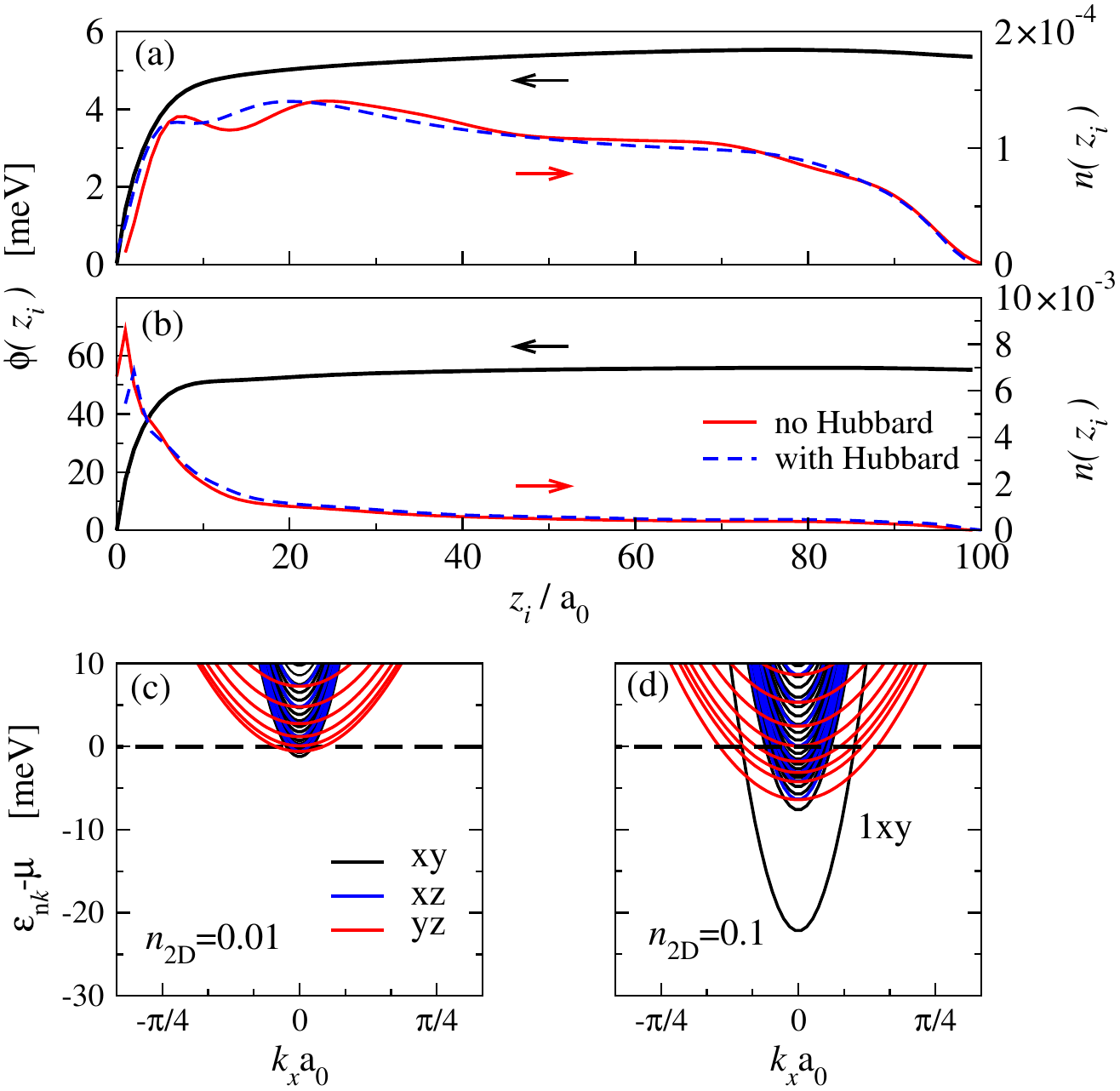}
\caption{The ideal  interface.  The confining potential and 3D charge density are plotted as functions of distance $z_i$ from the interface for (a) low ($n_\mathrm{2D} = 0.01$)  and (b) high ($n_\mathrm{2D} = 0.1$) electron density.  [$n_\mathrm{2D}$ and $n(z)$ are per 2D and 3D unit cell, respectively.]  Curves labeled ``with Hubbard'' include intra-atomic interactions, with $U_0 = 4$~eV, $U=2.4$~eV, and $J=0.8$~eV \cite{SM}; all other results have $U_0=U=J=0$.  The corresponding band structures (without Hubbard interactions) are shown in (c) and (d).  Unless otherwise stated, all results in this work are at $T=1$~K and for $L=100$ layers. }
\label{fig:noLifshitz}
\end{figure}

Figure~\ref{fig:noLifshitz} illustrates the absence of a Lifshitz transition in an ideal interface.  In this case, ${\bf \tilde D}$ is equal to the matrix ${\bf D}$ of stiffness constants for bulk STO, and $\tilde E_i = E_i$ is the total electric field in layer $i$.    Figures~\ref{fig:noLifshitz}(a) and \ref{fig:noLifshitz}(b) show the electron potential energy $\phi(z)$ and 3D density $n(z)$ for low and high values of $n_\mathrm{2D}$.   The key point is that the depth of the potential well confining the 2DEG is approximately proportional to $n_\mathrm{2D}$, and at low density is too shallow to create 2D bound states. Thus, while most of the charge lies within $\sim 20$ unit cells (8~nm) of the interface when $n_\mathrm{2D} = 0.1$, it spreads far into the substrate when $n_\mathrm{2D}=0.01$, with $n(z)$ decaying as a power law \cite{Atkinson:2017jt}.  (Densities are per 2D or 3D unit cell, as appropriate.) The electron confinement at high density is because  a large fraction of the 2DEG occupies a single 2D band (``1xy''), while the deconfinement at low density is because the electrons are shared amongst a nearly continuous spectrum of subbands.   That multiple bands are occupied at low $n_\mathrm{2D}$ is inconsistent with experiments.

The situation is not substantively changed when Hubbard-like interactions are included, and indeed the charge distributions with and without Hubbard interactions are nearly the same [Fig.~\ref{fig:noLifshitz}(a) and Fig.~\ref{fig:noLifshitz}(b)].

The shortcomings of the ideal interface are corrected if strain is included.
For a layer-dependent strain $\eta_i$,
\begin{equation}
\tilde D_{ij} = D_{ij} - 2\delta_{i,j} g_{11} \eta_i,\qquad 
\tilde E_i =  E_i +  f_{11}  \left. \frac{\partial \eta}{\partial z} \right|_{z=z_i},
\label{eq:Dmod1}
\end{equation}
with $g_{11}$ and $f_{11}$ the coupling constants, and $\delta_{i,j}$ the Kronecker delta-function.
We adopt an empirical expression that qualitatively fits experimental measurements of the strain profile \cite{Willmott:2007ip,Lee:2016dj},
$\eta_i = \eta_1 \exp[{-(z_i/d)^4}]$,
with $\eta_1 \sim 0.01$--0.03  the strain at the top STO layer and $d=4a_0$.  On general grounds, a piezoelectric term, $-e(z_i)\eta_i$ should also be added to $\tilde E_i$, where the coupling constant $e(z_i)$ vanishes away from the interface \cite{Yudin:2013}.   Because $e(z_i)$ is unknown, and because it plays the same qualitative role as flexoelectricity, surface piezoelectricity will not be considered explicitly.

Key to Eq.~(\ref{eq:Dmod1}) is that, because $\partial_z \eta$ is negative, the effective field $\tilde E_i$ can be negative when the electric field $E_i$ is sufficiently small, provided $f_{11}>0$.  In this case, the polarization at the interface will point {\em oppositely} to the external field, and towards the interface.  This allows for a switchable polarization as a function of gate voltage.  

Unless otherwise stated, we adopt the bulk value $g_{11}=0.118 \epsilon_0^{-1}$, with $\epsilon_0$ the permittivity of free space \cite{SM,Uwe:1976fj}.  The appropriate value of $f_{11}$ is harder to asses \cite{Hong:2011hc,Hong:2010kx,Zubko:2007kh,*Zubko:2008gh}, first because it is difficult to measure, even in bulk \cite{Zubko:2007kh,*Zubko:2008gh}, second because surface corrections should be comparable to the bulk value \cite{Yudin:2013}, and third because screening by the 2DEG modifies the lattice response to a longitudinal strain gradient.  $f_{11}$ is thus the only unknown parameter in our model.  One empirical guide is a recent observation that the STO surface polarization in ungated samples points towards the interface \cite{Lee:2016dj}, implying $f_{11} > 0$.  We then take   $f_{11}$ of order  a few~V \cite{Zubko:2013bt}, which is typical for perovskites.

\begin{figure}[tb]
\includegraphics[width=\columnwidth]{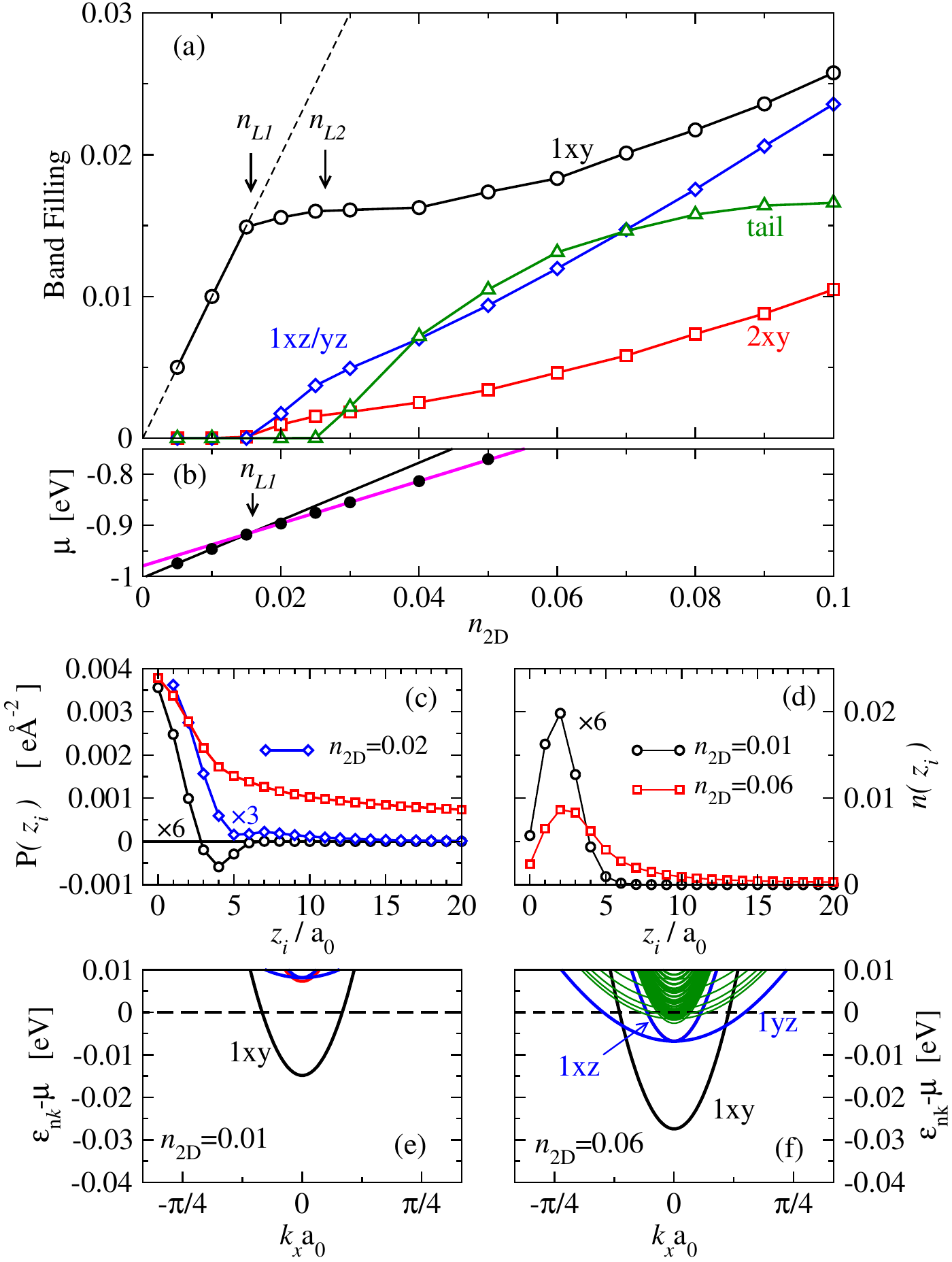}
\caption{Effects of strain on the 2DEG.  (a)  The fillings of the lowest four bands are shown individually, along with the cumulative filling of the remaining bands  (``tail'') and the total electron density (dashed line), as a function of $n_\mathrm{2D}$.  Bands are labeled $j\alpha$, with $j$ the band index and $\alpha$ the band symmetry.  There are two Lifshitz transitions,  at $n_{L1}$ and $n_{L2}$.  (b) The chemical potential is plotted versus $n_\mathrm{2D}$ (symbols).  Black and magenta lines are linear fits to the regions $n_\mathrm{2D} < n_{L1}$ and $n_{L1}< n_\mathrm{2D} < n_{L2}$, highlighting the discontinuity in slope at $n_{L1}$.  Any discontinuity at $n_{L2}$ is too weak to see.
The (c) polarization and (d) electron density are plotted for the top 20 STO layers for $n_\mathrm{2D}$ below and above the transitions.  A positive polarization points away from the interface.   The corresponding band structures are shown in (e) and (f).  In (f), the 2xy band is obscured by the 1yz band.  The tail bands in (a) correspond to the dense spectrum of unlabeled bands in (f).
Here, $f_{11} = 2$~V and $\eta_1 = 0.02$. }
\label{fig:straingradient1}
\end{figure}

Figure~\ref{fig:straingradient1}(a) shows the filling of the four lowest bands as a function of $n_\mathrm{2D}$, along with the total filling of the remaining bands making up the tails of the charge distribution.  There is a clear Lifshitz transition at $n_{L1} \approx 0.0154$  at which the slope of $n_{1xy}$ (the filling of the 1xy band) is discontinuous:   all electrons reside in the 1xy band when $n_\mathrm{2D} < n_{L1}$, and $n_{1xy}$ is nearly constant over an extended region when $n_\mathrm{2D} > n_{L1}$.  That such a transition emerges naturally from the model without any parameter tuning, and that the predicted transition lies close to the experimental value of $n_L$, is remarkable.

There is a second transition at $n_{L2}\approx 0.025$; this has a small effect on $n_{1xy}$, and is primarily a redistribution of electrons from the 1xz/yz and 2xy bands into the quasi-3D tails.  Although no more than 20\% of the charge occupies the tails, their 2D density of states is orders-of-magnitude larger than that of the 1xy band, which makes them a highly effective charge reservoir.

The transition at $n_{L1}$ is sharp, suggesting that there are two competing ground states, and indeed the first-order nature of the transition is illustrated by Fig.~\ref{fig:straingradient1}(b), which shows a discontinuity in the inverse isothermal compressibility $\kappa_T^{-1} \propto d\mu/d n_\mathrm{2D}$.  There is, presumably, a second discontinuity at $n_{L2}$; however it is too small to resolve in our data.

To characterize the  states on either side of the transition, we show in Fig.~\ref{fig:straingradient1}(c) and Fig.~\ref{fig:straingradient1}(d) the lattice polarization and electron distribution, respectively.  The main feature of this figure is that for $n_\mathrm{2D} < n_{L1}$, there is a  thin layer of negative polarization that extends over the region $3a_0 \leq z_i \leq 6a_0$, where the strain gradient is largest.  In this region, the polarization points towards the interface, and opposite to the electric field.   This enhances the confining potential,  and pulls electrons into a narrow 2D layer sandwiched between the interface and the region of negative polarization [Fig.~\ref{fig:straingradient1}(c)].   When $n_\mathrm{2D}>n_{L1}$, the electric field is strong enough to overcome flexoelectric effects, and the polarization switches abruptly to be positive everywhere.   In this regime, the electrons move away from the interface and begin to occupy other bands.  When $n_\mathrm{2D}>n_{L2}$, the tail states become occupied, and the polarization develops a power-law decay into the substrate \cite{Atkinson:2017jt}. The corresponding band structures show a transition from a single occupied $d_{xy}$ band at low density, to multiple occupied bands at high density [Figs.~\ref{fig:straingradient1}(e) and \ref{fig:straingradient1}(f)].  
 
\begin{figure}
\includegraphics[width=\columnwidth]{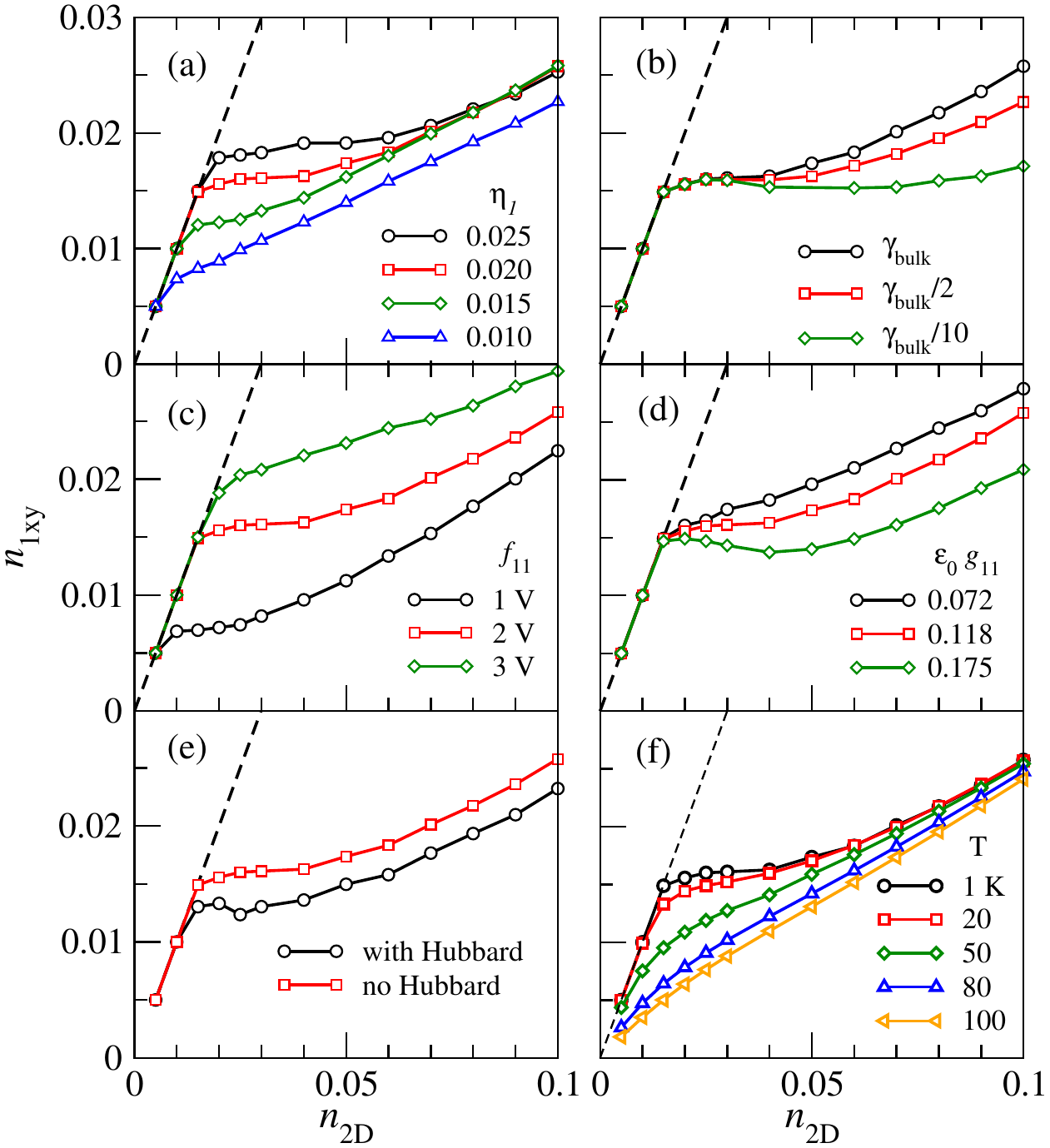}
\caption{Factors influencing the band filling.  The 1xy band filling is shown for different values of (a) the interfacial strain, (b) the coefficient of the quartic term in Eq.~(\ref{eq:D}), (c) the flexoelectric  constant, and (d) the electrostrictive constant.  In (e), cases with and without Hubbard-like interactions are compared.  The ``with Hubbard'' calculations assume $U_0 = 4$~eV, $U=2.4$~eV, and $J=0.8$~eV.  In (f), $n_{1xy}$ is shown for different temperatures.  Default model parameters are as in Fig.~\ref{fig:straingradient1} and $\gamma_\mathrm{bulk} = 2750$ eV\AA$^5e^{-4}$ is the estimate for $\gamma$ in bulk samples reported in \cite{Raslan:2017gh}.
 }
\label{fig:straingradient4}
\end{figure}

Figure~\ref{fig:straingradient4} shows   different factors affecting  $n_{1xy}$.  Both the interfacial strain [Fig.~\ref{fig:straingradient4}(a)] and the flexoelectric coupling constant [Fig.~\ref{fig:straingradient4}(c)] have a strong impact on the critical doping, consistent with our earlier assertion that  the polarization switches direction at the value of $n_{L1}$ where the electric field exceeds the flexoelectric term in Eq.~(\ref{eq:Dmod1}).  That  $n_{L1}$ depends on strain provides a natural explanation for the observed variability between samples of the critical doping.

Conversely, neither the quartic coefficient $\gamma$ [Fig.~\ref{fig:straingradient4}(b)] nor the electrostrictive coupling constant $g_{11}$ [Fig.~\ref{fig:straingradient4}(d)]  has a significant effect on $n_{L1}$.  Rather, they determine the behavior of $n_{1xy}$ on the high-density side of the transition.  The quartic term, which progressively reduces  the dielectric screening as $n_\mathrm{2D}$ grows, is responsible for an upturn in $n_{1xy}$ at large $n_\mathrm{2D}$, while    $g_{11}$ principally affects the slope  on the high density side of the Lifshitz transition, with large values of $g_{11}$ actually leading to a decline in $n_{1xy}$ with increasing $n_\mathrm{2D}$.  

The slope of $n_{1xy}$ above the Lifshitz transition is also strongly affected by the doping-dependence of the interfacial strain.  In Fig.~\ref{fig:straingradient4}(a), the strain is fixed for each curve; however, the strain is not constant in real interfaces and, for example, shrinks with increasing LAO thickness \cite{Lee:2016dj}.  The effects of gating have not been reported; however, one may infer from Fig.~\ref{fig:straingradient4}(a) that if the strain were to relax with increasing $n_\mathrm{2D}$, the slope above the transition would be negative, as found in some experiments \cite{Smink:2017cc,Niu:2017ih,Smink:2018tu}.

Figure~\ref{fig:straingradient4}(e) shows that the inclusion of intra-atomic interactions has a small quantitative effect on $n_{1xy}$, but does not change qualitative features of the transition.   This is consistent with our general finding that electron densities are too small for Hubbard-like interactions to have a significant effect. 

Figure~\ref{fig:straingradient4}(f) shows the evolution of $n_{1xy}$ with $T$.  While temperature appears directly in the calculation of the band fillings, the most important effect is in the stiffness coefficients $D_{ij}$, which reflect STO's strongly $T$-dependent dielectric function \cite{SM}.  In Fig.~\ref{fig:straingradient4}(f), the transition changes very little for $T\lesssim 20$~K, over which range $D_{ij}$ is nearly constant, and then is gradually wiped out as the temperature is further raised.  The mechanism for the wipeout is straightforward; as $T$ increases, the lattice stiffens (i.e.\ the permittivity decreases), and the spontaneously polarized layer near the interface disappears.

The predictions made here can be directly tested by atomic-resolution probes that can resolve the interface polarization. Notably, Lee {\em et al.} \cite{Lee:2016dj} observed a ``head-to-head'' arrangement of LAO and STO polarizations in ungated samples, which they associated with the interfacial doping mechanism. In their measurements, the reversed polarization on the STO side of the interface extends $\sim 5$ unit cells into the substrate, similar to what we find.  In this regard, it would be extremely interesting to see whether the measured polarization can be switched  by an external gate voltage, and to determine whether this correlates with a Lifshitz transition.
  
In summary, we have argued that LAO/STO interfaces can be thought of as a metallic system with ferroelectric characteristics arising from interfacial strains.  This is reminiscent of an earlier proposal that the 2DEG at interfaces beween LAO and a bulk ferroelectric, BaTiO$_3$, should exhibit a switchable metallic state \cite{Niranjan:2009ix}; in that case, the external electric field due to the LAO cap layer was ultimately shown suppress switchability \cite{Wang:2010cp}.  The current model differs in two key respects:  here, (i) switchability is a consequence of the competition between flexoelectric effects and the external field, and (ii) the switchable region extends over only a few unit cells.

We thank A.\ E.\ M.\ Smink for helpful comments.  We acknowledge support by the Natural Sciences and Engineering Research Council (NSERC) of Canada.   A.~R.\ was supported by an Ontario Graduate Scholarship.

%


\appendix
\section*{Supplemental Material for:  Possible Flexoelectric Origin of the Lifshitz Transition in LaAlO$_3$/SrTiO$_3$ Interfaces}

\subsection{Electronic Hamiltonian}
The electronic Hamiltonian is 
\begin{equation}
\label{H}
\hat H=\hat H_0 + \hat H_V + \hat H_U.
\end{equation}
In this expression,  $\hat H_0$ is the kinetic energy portion of the Hamiltonian, $\hat H_V$ describes the potential energy due to the long-range Coulomb interaction, obtained by solving the Poisson equation, while $\hat H_U$ contains short-range (intra-atomic) Coulomb contributions of the Hubbard type.

We adopt a layered geometry consistent with the structure of STO interfaces; we assume a (001) interface, such that there is translational invariance in the $x$-$y$ plane, and quantities such as the charge density, lattice polarization, and electrostatic potential depend only on the distance $z$ from the interface.  For this geometry,
\begin{equation}
\hat H_0=\sum_{ij\bk}\sum_{\alpha \sigma} c^\dagger_{\alpha \bk i \sigma} t^\alpha_{i j}({\bf k})  c_{\alpha \bk j \sigma},
\end{equation}
where $i$ and $j$ label TiO$_2$ layers,  and  ${\bf k}= (k_x,k_y)$ is the 2D wavevector describing the motion of electrons parallel to the interface.  The operator $c_{\alpha \bk j \sigma}$ annihilates an electron with orbital type $\alpha$, spin $\sigma$, and wavevector $\bk$ in layer $j$.  The matrix element $t^\alpha_{ij}({\bf k})$ is a hopping matrix element between orbitals of type $\alpha$.

There are two distinct hopping processes:  hopping between neighboring orbitals of type $\alpha$ is large if the electron moves in the plane parallel to $\alpha$ (e.g.\ the $x$-$y$ plane for $d_{xy}$ orbitals), and small if it moves perpendicular to $\alpha$.  The matrix elements are denoted by $t^{\|}$ and $t^{\perp}$, respectively, and  values are given in Table~\ref{table:1}.  We thus have
\begin{eqnarray}
	t^{xy}_{ij}({\bf k}) &=&  -2 (t^\parallel c_x + t^\parallel c_y  )  \delta_{i,j} - t^\perp \delta_{\langle i,j \rangle}, \\
	t^{xz}_{ij}({\bf k}) &=&  -2 (t^\parallel c_x + t^\perp c_y ) \delta_{i,j} - t^\|\delta_{\langle i,j \rangle}, \\
        t^{yz}_{ij}({\bf k})&=& -2 ( t^\perp c_x + t^\parallel c_y ) \delta_{i,j} - t^\|\delta_{\langle i,j \rangle},
\end{eqnarray}
where $c_{n} \equiv \cos(k_n a_0)$ with $a_0$ the lattice constant, and $\delta_{\langle i,j\rangle}$ is one if $i$ and $j$ are nearest-neighbor planes, and zero otherwise.

\begin{table} 
 \begin{tabular}{l | r}
  \hline
  \multicolumn{2}{c}{Model parameters} \\
  \hline
 $t^\parallel $& 0.236 eV\\
  $t^\perp$ & 0.035 eV\\
 $a_0$ & 3.9 \AA \\
 $D_0$ & $5.911 \times 10^{-3} \epsilon_0^{-1}$\\ 
 $D_1$ & $1.200 \times 10^{-3}  \epsilon_0^{-1}$\\ 
 $D_2(T=0)$ & $0.312 \times 10^{-3}  \epsilon_0^{-1}$\\ 
  $\alpha_1$ & $1.15a_0$ \\
 $\alpha_2$ & $5a_0$  \\
$\gamma_\mathrm{bulk}$ & 2750 eV\AA$^5$e$^{-4}$\\
 $\epsilon_\infty$ & $5.5\epsilon_0$\\
 $g_{11}$ & $0.118 \epsilon_0^{-1}$ \\
 $f_{11}$ & 1--3 V \\
   \hline
\end{tabular}
  \caption{Model parameters used in our calculations.    
  Hopping matrix elements $t^\|$ and $t^\perp$ are from tight-binding fits \cite{Khalsa}  to Shubnikov-de Haas measurements in bulk STO \cite{Allen}.  Dielectric parameters $D_0$, $D_1$, $D_2$, $\alpha_1$,  $\alpha_2$, and $\gamma_\mathrm{bulk}$ are obtained from fits to the dielectric function.\cite{Amany,Dec}, while $g_{11}$ is obtained from Ref.~\cite{Uwe}.
  $\epsilon_0$ is the permittivity of free space and $-e$ is the electron charge. }
  \label{table:1}
\end{table}

\subsection{Coulomb Potential}
The long-range Coulomb term is
\begin{equation}
\hat H_V = \sum_{i \alpha \sigma}  \phi_i \hat n_{\alpha i \sigma},
\end{equation}
where $\hat n_{\alpha i \sigma}=  N_{2D}^{-1} \sum_\bk c^\dagger_{\alpha \bk i \sigma}   c_{\alpha \bk i \sigma}$ is the electron number operator for orbital type $\alpha$ and spin $\sigma$ in  layer $i$, with $N_{2D}$ the number of 2D unit cells.
$\hat H_V$ contains a potential energy $\phi_i \equiv \phi(z_i)$ with contributions from three distinct interactions: electron-electron interactions between free carriers in the STO, electron-lattice interactions between free carriers and the STO polarization, and interactions between free carriers and  the LAO surface charge.  

The electrostatic potential energy (in SI units) is obtained by solving Poisson's equation,
\begin{equation}
\epsilon_\infty \nabla \cdot {\bf E}(z) = \rho (z) - \nabla \cdot {\bf P}(z),
\label{eq:gauss}
\end{equation}
where ${\bf E}(z)$, ${\bf P}(z)$, and $\rho(z)$ are, respectively, the electric field, the lattice polarization, and the charge density, and $\epsilon_\infty$ is the optical dielectric constant.   The charge density includes contributions from both the free carriers in the STO substrate and LAO surface charges.  In the planar geometry, the polarization and electric field vectors are parallel to the $z$-axis, with ${\bf E}(z) = E(z) {\bf \hat z}$, ${\bf P}(z) = P(z) {\bf \hat z}$.  Then, the potential energy is obtained from
\begin{equation}
\phi(z) = \phi(0) +e \int_{0}^{z} dz' \, E(z').
\label{eq:phi}
\end{equation}
  Integrating Eq.~(\ref{eq:gauss}) from a point outside the LAO surface (where $E$ and $P$ vanish) to layer $i$ inside the STO substrate yields
\begin{equation}
\epsilon_\infty E_i = e n_\mathrm{2D}  -\frac{e}{a_0^2} \sum_{j<i} \sum_{\alpha \sigma} n_{\alpha j \sigma} -P_i.
\label{eq:Efield}
\end{equation}
Equation~(\ref{eq:phi}) then gives the potential energy,
\begin{equation}
\phi_i =  \frac{e^2}{\epsilon_\infty}\left [ n_\mathrm{2D}  z_i  -  \sum_{\stackrel{j<i}{\alpha \sigma}} \frac{n_{\alpha j\sigma}}{a_0^2} (z_i-z_j) \right ]
- \frac{ea_0}{\epsilon_\infty} \sum_{j\leq i} P_j,
\end{equation}
where $\phi(0)$ is set to zero.

We solve a discrete version of these equations, in which $E(z) \rightarrow E_i$, $P(z) \rightarrow P_i$, and 
\begin{equation}
\rho(z) = en_\mathrm{2D} \delta(z-z_\mathrm{surf}) - \frac{e}{a_0^2} \sum_{\alpha i \sigma} n_{\alpha i\sigma}\delta(z-z_i) .
\end{equation}
Here, $z_\mathrm{surf}$ is the location of the LAO surface, $z_i = (i-1) a_0$ is the location of the $i$th layer measured relative to the interface, and 
\begin{equation}
n_{\alpha i\sigma} =  \langle \hat n_{\alpha i \sigma}\rangle.
\end{equation}
is the  occupation of a single orbital $(\alpha,\sigma)$ in layer $i$. 

The short-range Coulomb interaction on the Ti sites takes the form  \cite{Oles},
\begin{equation}
\hat H_U = \sum_{ \alpha\sigma i} \left [ U_0 n_{\alpha i \sigmab} + \sum_{\beta\neq \alpha, \sigma'}
\left ( U - J\delta_{\sigma,\sigma'} \right ) n_{\beta i \sigma'}
\right ]  \hat n_{\alpha i \sigma},
\label{eq:HU}
\end{equation}
where $\sigmab = -\sigma$ and with the constraint 
\begin{equation}
U_0 = U + 2J.
\label{eq:constraint}
\end{equation}  In this expression, $U_0$ is the intra-orbital Hubbard interaction, and $U$ and $J$ are inter-orbital Hartree and exchange interactions.  We constrain our calculations to eliminate the possibility of ferromagnetism, and require that $n_{\alpha i \downarrow} = n_{\alpha i \uparrow}$. 

From the form of $\hat H_U$  [Eq.~(\ref{eq:HU})], it is clear that minimization of the intra-orbital Coulomb energy favors a spreading-out of charge between orbitals if $U_0$ and $J$ are large, and the collapse of charge into a single orbital type if $U$ is large.  Indeed, we find that our calculations favor occupation of $d_{xy}$ orbitals over $d_{xz}$ and $d_{yz}$ orbitals when $J < U_0/5$, and favor occupation of multiple orbital types when $J > U_0/5$.  The representative example shown in the manuscript takes $U_0 = 4$~eV, and $J = 0.8$~eV (so $J = U_0/5$) and from  the constraint (\ref{eq:constraint}), we obtain $U = 2.4$~eV.    We have explored other $J$-values, and find that while details of the band structure change, our basic conclusions do not.

\subsection{Dielectric Model at Zero Temperature}
\label{sec:dielectric}
Dielectric screening comes from a soft optical phonon mode with a large dipole moment.  
The Landau-Devonshire free energy  for this mode is  
\begin{equation}
{\cal U}=\frac{1}{2}\sum_{i,j} P_i {\tilde D}_{ij}P_j -  \sum_{i} \tilde E_i P_i +\mbox{quartic term}
\label{eq:Dapp}
\end{equation}
where $i$ and $j$ are layer indices, and $\tilde E_i$ and $\tilde D_{ij}$ are linear and quadratic coefficients of the free energy expansion. The quartic term is discussed below.

In bulk STO, $\tilde E_i$ is equal to the electric field $E_i$; adjacent to the interface, inversion symmetry is broken and $\tilde E_i$ picks up additive corrections that extend over the first few STO layers.  Similarly, the coefficients $\tilde D_{ij}$ equal their bulk values  $D_{ij}$ away from the interface, but may be different near the interface.    

Following our earlier work, we take the bulk coefficients
\begin{equation}
D_{ij} = \left \{ \begin{array}{ll}
D_0, & i=j \\
-D_1 e^{-z_{ij}^2/2\alpha_1^2} - D_2 e^{-z_{ij}^2/2\alpha_2^2}, & i\neq j \\
\end{array}\right .
\label{eq:Dij}
\end{equation}
where $z_{ij} = z_i -z_j$.  Values for $D_0$, $D_1$, $D_2$, $\alpha_1$, and $\alpha_2$ that are valid at low $T$ are given in Table~\ref{table:1}.    

For a given electric field, the polarization $P_\ell$ is obtained by setting $\partial {\cal U}/\partial P_\ell =0$.  In a bulk crystal with uniform electric field $E$ and polarization $P$, this yields $P = E/D_{\bq=0}$ where $D_{\bq=0} = \sum_j D_{ij}$.  The bulk dielectric constant is then
\begin{equation}
\epsilon = \epsilon_\infty + \frac{\partial P}{\partial E} \approx D_{\bq=0}^{-1}.
\label{eq:eps}
\end{equation}
The final approximate equality arises because $\epsilon_\infty = 5.5 \epsilon_0$ while $D_{\bq=0}^{-1} \sim 10^4 \epsilon_0$ at low $T$ ($\epsilon_0$ is the permittivity of free space).  In this dielectric model, the large value of $\epsilon$ comes from a near cancellation of the local ($i=j$) and nonlocal ($i\neq j$) contributions to $D_{\bq=0}$.

Electrostrictive coupling through the term proportional to $g_{11}$ modifies the matrix ${\bf \tilde D}$ such that it has one or two negative eigenvalues, depending on the magnitude of the strain.  In this instance, the quartic terms in Eq.~(\ref{eq:Dapp}) are required to stabilize the polarization.  We find that the numerics are most easily controlled if we work in a basis in which ${\bf \tilde D}$ is diagonal.  Letting $\Lambda_n$ and ${\bf S}$ be the eigenvalues and the matrix of eigenvectors of ${\bf\tilde D}$, we  make the ansatz
\begin{equation}
{\cal U}= \sum_{n}   \left [ \frac{1}{2}{\Lambda}_{n} {\cal P}_n^2 -  {\cal E}_n {\cal P}_n + \frac{\gamma}{4} {\cal P}_n^4 \right ],
\label{eq:D2}
\end{equation}
where ${\cal E}_n = \sum_i \tilde E_i S_{in}$ and ${\cal P}_n = \sum_i P_i S_{in}$.  The first two terms in Eq.~(\ref{eq:D2}) are formally equivalent to Eq.~(\ref{eq:Dapp}), while the final term is an ansatz.  The advantage of Eq~(\ref{eq:D2}) is that it is diagonal in the mode index $n$, and one can minimize each term in the sum analytically.   Once ${\cal P}_n$ is known, then $P_i$ is obtained from $P_i = \sum_n S_{in} {\cal P}_n$.

At low electron densities, ${\cal E}_n$ is sufficiently weak that the quartic term is negligible provided $\Lambda_n$ is positive and not too small; then, ${\cal P}_n = {\cal E}_n/\Lambda_n$, to a good approximation.  For the one or two eigenmodes where $\Lambda_n$ is close to zero or negative,  $\gamma$ cannot be neglected and the cubic equation obtained from setting 
\begin{equation}
\partial{\cal U}/\partial{\cal P}_n = \Lambda_n {\cal P}_n - {\cal E}_n + \gamma {\cal P}_n^3 = 0
\label{eq:cubic}
\end{equation}
must be solved, with the solution that minimizes the energy of that eigenmode being selected. 

\subsection{Dielectric Model at Nonzero Temperature}
The temperature dependence of the dielectric function comes through the matrix elements $D_{ij}$, and our approach follows  Raslan {\em et al.} \cite{Amany}.  The $T$-dependence is obtained in two steps. First, the measured dielectric susceptibility \cite{Dec}.
$\chi(T)$ is fitted to an empirical formula
\begin{equation}
\chi_{\bq=0}(T) = \left(\frac{T_0}{T_s \coth(T_s/T)} \right )^\xi,
\label{eq:chi}
\end{equation}
where $T_s=15$~K is the saturation temperature below which $\chi(T)$ becomes constant, and $T_0=1.46\times 10^4$~K and $\xi = 1.45$ are  fitting parameters.  Next, using
\begin{equation}
\chi_{\bq=0}(T) = \frac{1}{\epsilon_0 D_{\bq=0} },
\end{equation}
allows us to obtain the temperature dependent parameters in $D_{ij}$.  We write
\begin{eqnarray}
D_{\bq=0} &=& \sum_{j} D_{ij} \nonumber \\
&\approx& D_0 - D_1\left (1-\frac{\sqrt{2\pi}\alpha_1}{a_0}\right  ) 
- D_2 \left (1-\frac{\sqrt{2\pi}\alpha_2}{a_0}\right  ). \nonumber \\
\label{eq:Dq0}
\end{eqnarray}
Comparing Eq.~(\ref{eq:Dq0}) and Eq.~(\ref{eq:chi}) allows us to determine the $T$-dependence of our model parameters.  Following Ref.~\cite{Khalsa},
we make the ansatz that $D_2$ is $T$-dependent, while $D_0$ and $D_1$ are constant.

\subsection{Numerical Solution of the Model}
\label{sec:numerics}
In most cases, the model can be solved using a straightforward iterative procedure to obtain self-consistent values for the polarization $P_i$ and the electron density $n_i$.  Given an input potential energy $\phi_i$, we calculate the charge density $n_{\alpha i \sigma}$ by diagonalizing $\hat H$ and the polarization $P_i$ by minimizing ${\cal U}$.  The charge density and polarization are then used to generate an updated potential energy.  The cycle is repeated until the input and output potentials are the same.  This cycle is unstable to charge-sloshing, and we therefore use Anderson mixing to stabilize the iterative process.\cite{Eyert}

In some of the cases, the iterative scheme described above is unstable because the flexoelectric contribution to $\tilde E$ can lead to a rapid switching of the polarization direction from one iteration to the next.  To stabilize the numerics, we rearrange Eq.~(\ref{eq:cubic}) such that the depolarizing fields are explicitly grouped with $\Lambda_n$.    Thus, from Eq.~(\ref{eq:Efield}) it is possible to write $E_i = -P_i/\epsilon_\infty + E^\mathrm{other}_i$, from which it follows that Eq.~(\ref{eq:cubic}) is 
\begin{equation}
 ( \Lambda_n + \epsilon_\infty^{-1} ) {\cal P}_n - {\cal E}_n^\mathrm{other} + \gamma {\cal P}_n^3 = 0.
\end{equation} 
In our calculations, $\Lambda_n > -\epsilon_\infty^{-1}$ and this rearrangement of terms thus helps stabilize the iterative cycle.


\begin{thebibliography}{59}%
\makeatletter
\providecommand \@ifxundefined [1]{%
 \@ifx{#1\undefined}
}%
\providecommand \@ifnum [1]{%
 \ifnum #1\expandafter \@firstoftwo
 \else \expandafter \@secondoftwo
 \fi
}%
\providecommand \@ifx [1]{%
 \ifx #1\expandafter \@firstoftwo
 \else \expandafter \@secondoftwo
 \fi
}%
\providecommand \natexlab [1]{#1}%
\providecommand \enquote  [1]{``#1''}%
\providecommand \bibnamefont  [1]{#1}%
\providecommand \bibfnamefont [1]{#1}%
\providecommand \citenamefont [1]{#1}%
\providecommand \href@noop [0]{\@secondoftwo}%
\providecommand \href [0]{\begingroup \@sanitize@url \@href}%
\providecommand \@href[1]{\@@startlink{#1}\@@href}%
\providecommand \@@href[1]{\endgroup#1\@@endlink}%
\providecommand \@sanitize@url [0]{\catcode `\\12\catcode `\$12\catcode
  `\&12\catcode `\#12\catcode `\^12\catcode `\_12\catcode `\%12\relax}%
\providecommand \@@startlink[1]{}%
\providecommand \@@endlink[0]{}%
\providecommand \url  [0]{\begingroup\@sanitize@url \@url }%
\providecommand \@url [1]{\endgroup\@href {#1}{\urlprefix }}%
\providecommand \urlprefix  [0]{URL }%
\providecommand \Eprint [0]{\href }%
\providecommand \doibase [0]{http://dx.doi.org/}%
\providecommand \selectlanguage [0]{\@gobble}%
\providecommand \bibinfo  [0]{\@secondoftwo}%
\providecommand \bibfield  [0]{\@secondoftwo}%
\providecommand \translation [1]{[#1]}%
\providecommand \BibitemOpen [0]{}%
\providecommand \bibitemStop [0]{}%
\providecommand \bibitemNoStop [0]{.\EOS\space}%
\providecommand \EOS [0]{\spacefactor3000\relax}%
\providecommand \BibitemShut  [1]{\csname bibitem#1\endcsname}%
\let\auto@bib@innerbib\@empty
\bibitem [{\citenamefont {Ohtomo}\ and\ \citenamefont
  {Hwang}(2004)}]{Ohtomo:2004hm}%
  \BibitemOpen
  \bibfield  {author} {\bibinfo {author} {\bibfnamefont {A.}~\bibnamefont
  {Ohtomo}}\ and\ \bibinfo {author} {\bibfnamefont {H.~Y.}\ \bibnamefont
  {Hwang}},\ }\bibfield  {title} {\enquote {\bibinfo {title} {{A high-mobility
  electron gas at the LaAlO$_3$/SrTiO$_3$ heterointerface}},}\ }\href@noop {}
  {\bibfield  {journal} {\bibinfo  {journal} {Nature Commun.}\ }\textbf
  {\bibinfo {volume} {427}},\ \bibinfo {pages} {423--426} (\bibinfo {year}
  {2004})}\BibitemShut {NoStop}%
\bibitem [{\citenamefont {Thiel}\ \emph {et~al.}(2006)\citenamefont {Thiel},
  \citenamefont {Hammerl}, \citenamefont {Schmehl}, \citenamefont {Schneider},\
  and\ \citenamefont {Mannhart}}]{Thiel:2006eo}%
  \BibitemOpen
  \bibfield  {author} {\bibinfo {author} {\bibfnamefont {S.}~\bibnamefont
  {Thiel}}, \bibinfo {author} {\bibfnamefont {G.}~\bibnamefont {Hammerl}},
  \bibinfo {author} {\bibfnamefont {A.}~\bibnamefont {Schmehl}}, \bibinfo
  {author} {\bibfnamefont {C.~W.}\ \bibnamefont {Schneider}}, \ and\ \bibinfo
  {author} {\bibfnamefont {J.}~\bibnamefont {Mannhart}},\ }\bibfield  {title}
  {\enquote {\bibinfo {title} {{Tunable Quasi-Two-Dimensional Electron Gases in
  Oxide Heterostructures}},}\ }\href@noop {} {\bibfield  {journal} {\bibinfo
  {journal} {Science}\ }\textbf {\bibinfo {volume} {313}},\ \bibinfo {pages}
  {1942--1945} (\bibinfo {year} {2006})}\BibitemShut {NoStop}%
\bibitem [{\citenamefont {Caviglia}\ \emph {et~al.}(2008)\citenamefont
  {Caviglia}, \citenamefont {Gariglio}, \citenamefont {Reyren}, \citenamefont
  {Jaccard}, \citenamefont {Schneider}, \citenamefont {Gabay}, \citenamefont
  {Thiel}, \citenamefont {Hammerl}, \citenamefont {Triscone},\ and\
  \citenamefont {Mannhart}}]{Caviglia:2008uh}%
  \BibitemOpen
  \bibfield  {author} {\bibinfo {author} {\bibfnamefont {A.~D.}\ \bibnamefont
  {Caviglia}}, \bibinfo {author} {\bibfnamefont {S.}~\bibnamefont {Gariglio}},
  \bibinfo {author} {\bibfnamefont {N.}~\bibnamefont {Reyren}}, \bibinfo
  {author} {\bibfnamefont {D.}~\bibnamefont {Jaccard}}, \bibinfo {author}
  {\bibfnamefont {T.}~\bibnamefont {Schneider}}, \bibinfo {author}
  {\bibfnamefont {M.}~\bibnamefont {Gabay}}, \bibinfo {author} {\bibfnamefont
  {S.}~\bibnamefont {Thiel}}, \bibinfo {author} {\bibfnamefont
  {G.}~\bibnamefont {Hammerl}}, \bibinfo {author} {\bibfnamefont
  {J.}~\bibnamefont {Triscone}}, \ and\ \bibinfo {author} {\bibfnamefont
  {J.~M.}\ \bibnamefont {Mannhart}},\ }\bibfield  {title} {\enquote {\bibinfo
  {title} {{Electric field control of the LaAlO$_3$/SrTiO$_3$ interface ground
  state}},}\ }\href@noop {} {\bibfield  {journal} {\bibinfo  {journal}
  {Nature}\ }\textbf {\bibinfo {volume} {456}},\ \bibinfo {pages} {624--627}
  (\bibinfo {year} {2008})}\BibitemShut {NoStop}%
\bibitem [{\citenamefont {Dikin}\ \emph {et~al.}(2011)\citenamefont {Dikin},
  \citenamefont {Mehta}, \citenamefont {Bark}, \citenamefont {Folkman},
  \citenamefont {Eom},\ and\ \citenamefont {Chandrasekhar}}]{Dikin:2011gl}%
  \BibitemOpen
  \bibfield  {author} {\bibinfo {author} {\bibfnamefont {D.~A.}\ \bibnamefont
  {Dikin}}, \bibinfo {author} {\bibfnamefont {M.}~\bibnamefont {Mehta}},
  \bibinfo {author} {\bibfnamefont {C.~W.}\ \bibnamefont {Bark}}, \bibinfo
  {author} {\bibfnamefont {C.~M.}\ \bibnamefont {Folkman}}, \bibinfo {author}
  {\bibfnamefont {C.~B.}\ \bibnamefont {Eom}}, \ and\ \bibinfo {author}
  {\bibfnamefont {V.}~\bibnamefont {Chandrasekhar}},\ }\bibfield  {title}
  {\enquote {\bibinfo {title} {{Coexistence of Superconductivity and
  Ferromagnetism in Two Dimensions}},}\ }\href@noop {} {\bibfield  {journal}
  {\bibinfo  {journal} {Phys. Rev. Lett.}\ }\textbf {\bibinfo {volume} {107}},\
  \bibinfo {pages} {056802} (\bibinfo {year} {2011})}\BibitemShut {NoStop}%
\bibitem [{\citenamefont {Biscaras}\ \emph {et~al.}(2012)\citenamefont
  {Biscaras}, \citenamefont {Bergeal}, \citenamefont {Hurand}, \citenamefont
  {Grossetete}, \citenamefont {Rastogi}, \citenamefont {Budhani}, \citenamefont
  {LeBoeuf}, \citenamefont {Proust},\ and\ \citenamefont
  {Lesueur}}]{Biscaras:2012vd}%
  \BibitemOpen
  \bibfield  {author} {\bibinfo {author} {\bibfnamefont {J.}~\bibnamefont
  {Biscaras}}, \bibinfo {author} {\bibfnamefont {N.}~\bibnamefont {Bergeal}},
  \bibinfo {author} {\bibfnamefont {S.}~\bibnamefont {Hurand}}, \bibinfo
  {author} {\bibfnamefont {C.}~\bibnamefont {Grossetete}}, \bibinfo {author}
  {\bibfnamefont {A.}~\bibnamefont {Rastogi}}, \bibinfo {author} {\bibfnamefont
  {R.~C.}\ \bibnamefont {Budhani}}, \bibinfo {author} {\bibfnamefont
  {D.}~\bibnamefont {LeBoeuf}}, \bibinfo {author} {\bibfnamefont
  {C.}~\bibnamefont {Proust}}, \ and\ \bibinfo {author} {\bibfnamefont
  {J.}~\bibnamefont {Lesueur}},\ }\bibfield  {title} {\enquote {\bibinfo
  {title} {{Two-Dimensional Superconducting Phase in LaTiO$_{3}$/SrTiO$_{3}$
  Heterostructures Induced by High-Mobility Carrier Doping}},}\ }\href@noop {}
  {\bibfield  {journal} {\bibinfo  {journal} {Phys. Rev. Lett.}\ }\textbf
  {\bibinfo {volume} {108}} (\bibinfo {year} {2012})}\BibitemShut {NoStop}%
\bibitem [{\citenamefont {Maniv}\ \emph {et~al.}(2015)\citenamefont {Maniv},
  \citenamefont {Shalom}, \citenamefont {Ron}, \citenamefont {Mograbi},
  \citenamefont {Palevski}, \citenamefont {Goldstein},\ and\ \citenamefont
  {Dagan}}]{Maniv:2015cc}%
  \BibitemOpen
  \bibfield  {author} {\bibinfo {author} {\bibfnamefont {E.}~\bibnamefont
  {Maniv}}, \bibinfo {author} {\bibfnamefont {M.~Ben}\ \bibnamefont {Shalom}},
  \bibinfo {author} {\bibfnamefont {A.}~\bibnamefont {Ron}}, \bibinfo {author}
  {\bibfnamefont {M.}~\bibnamefont {Mograbi}}, \bibinfo {author} {\bibfnamefont
  {A.}~\bibnamefont {Palevski}}, \bibinfo {author} {\bibfnamefont
  {M.}~\bibnamefont {Goldstein}}, \ and\ \bibinfo {author} {\bibfnamefont
  {Y.}~\bibnamefont {Dagan}},\ }\bibfield  {title} {\enquote {\bibinfo {title}
  {{Strong correlations elucidate the electronic structure and phase diagram of
  LaAlO$_3$/SrTiO$_3$ interface}},}\ }\href@noop {} {\bibfield  {journal}
  {\bibinfo  {journal} {Nature Commun.}\ }\textbf {\bibinfo {volume} {6}},\
  \bibinfo {pages} {8239} (\bibinfo {year} {2015})}\BibitemShut {NoStop}%
\bibitem [{\citenamefont {Hurand}\ \emph {et~al.}(2015)\citenamefont {Hurand},
  \citenamefont {Jouan}, \citenamefont {Feuillet-Palma}, \citenamefont {Singh},
  \citenamefont {Biscaras}, \citenamefont {Lesne}, \citenamefont {Reyren},
  \citenamefont {Barth{\'e}l{\'e}my}, \citenamefont {Bibes}, \citenamefont
  {Villegas}, \citenamefont {Ulysse}, \citenamefont {Lafosse}, \citenamefont
  {Pannetier-Lecoeur}, \citenamefont {Caprara}, \citenamefont {Grilli},
  \citenamefont {Lesueur},\ and\ \citenamefont {Bergeal}}]{Hurand:2015cf}%
  \BibitemOpen
  \bibfield  {author} {\bibinfo {author} {\bibfnamefont {S.}~\bibnamefont
  {Hurand}}, \bibinfo {author} {\bibfnamefont {A.}~\bibnamefont {Jouan}},
  \bibinfo {author} {\bibfnamefont {C.}~\bibnamefont {Feuillet-Palma}},
  \bibinfo {author} {\bibfnamefont {G.}~\bibnamefont {Singh}}, \bibinfo
  {author} {\bibfnamefont {J.}~\bibnamefont {Biscaras}}, \bibinfo {author}
  {\bibfnamefont {E.}~\bibnamefont {Lesne}}, \bibinfo {author} {\bibfnamefont
  {N.}~\bibnamefont {Reyren}}, \bibinfo {author} {\bibfnamefont
  {A.}~\bibnamefont {Barth{\'e}l{\'e}my}}, \bibinfo {author} {\bibfnamefont
  {M.}~\bibnamefont {Bibes}}, \bibinfo {author} {\bibfnamefont {J.~E.}\
  \bibnamefont {Villegas}}, \bibinfo {author} {\bibfnamefont {C.}~\bibnamefont
  {Ulysse}}, \bibinfo {author} {\bibfnamefont {X.}~\bibnamefont {Lafosse}},
  \bibinfo {author} {\bibfnamefont {M.}~\bibnamefont {Pannetier-Lecoeur}},
  \bibinfo {author} {\bibfnamefont {S.}~\bibnamefont {Caprara}}, \bibinfo
  {author} {\bibfnamefont {M.}~\bibnamefont {Grilli}}, \bibinfo {author}
  {\bibfnamefont {J.}~\bibnamefont {Lesueur}}, \ and\ \bibinfo {author}
  {\bibfnamefont {N.}~\bibnamefont {Bergeal}},\ }\bibfield  {title} {\enquote
  {\bibinfo {title} {{Field-effect control of superconductivity and Rashba
  spin-orbit coupling in top-gated LaAlO$_3$/SrTiO$_3$ devices}},}\ }\href@noop
  {} {\bibfield  {journal} {\bibinfo  {journal} {Sci. Rep.}\ }\textbf {\bibinfo
  {volume} {5}},\ \bibinfo {pages} {12759} (\bibinfo {year}
  {2015})}\BibitemShut {NoStop}%
\bibitem [{\citenamefont {Caviglia}\ \emph {et~al.}(2010)\citenamefont
  {Caviglia}, \citenamefont {Gabay}, \citenamefont {Gariglio}, \citenamefont
  {Reyren}, \citenamefont {Cancellieri},\ and\ \citenamefont
  {Triscone}}]{Caviglia:2010jv}%
  \BibitemOpen
  \bibfield  {author} {\bibinfo {author} {\bibfnamefont {A.~D.}\ \bibnamefont
  {Caviglia}}, \bibinfo {author} {\bibfnamefont {M.}~\bibnamefont {Gabay}},
  \bibinfo {author} {\bibfnamefont {S.}~\bibnamefont {Gariglio}}, \bibinfo
  {author} {\bibfnamefont {N.}~\bibnamefont {Reyren}}, \bibinfo {author}
  {\bibfnamefont {C.}~\bibnamefont {Cancellieri}}, \ and\ \bibinfo {author}
  {\bibfnamefont {J.~M.}\ \bibnamefont {Triscone}},\ }\bibfield  {title}
  {\enquote {\bibinfo {title} {{Tunable Rashba Spin-Orbit Interaction at Oxide
  Interfaces}},}\ }\href@noop {} {\bibfield  {journal} {\bibinfo  {journal}
  {Phys. Rev. Lett.}\ }\textbf {\bibinfo {volume} {104}},\ \bibinfo {pages}
  {126803} (\bibinfo {year} {2010})}\BibitemShut {NoStop}%
\bibitem [{\citenamefont {Ben~Shalom}\ \emph {et~al.}(2010)\citenamefont
  {Ben~Shalom}, \citenamefont {Sachs}, \citenamefont {Rakhmilevitch},
  \citenamefont {Palevski},\ and\ \citenamefont {Dagan}}]{BenShalom:2010kv}%
  \BibitemOpen
  \bibfield  {author} {\bibinfo {author} {\bibfnamefont {M.}~\bibnamefont
  {Ben~Shalom}}, \bibinfo {author} {\bibfnamefont {M.}~\bibnamefont {Sachs}},
  \bibinfo {author} {\bibfnamefont {D.}~\bibnamefont {Rakhmilevitch}}, \bibinfo
  {author} {\bibfnamefont {A.}~\bibnamefont {Palevski}}, \ and\ \bibinfo
  {author} {\bibfnamefont {Y.}~\bibnamefont {Dagan}},\ }\bibfield  {title}
  {\enquote {\bibinfo {title} {{Tuning Spin-Orbit Coupling and
  Superconductivity at the SrTiO$_{3}$/LaAlO$_{3}$ Interface: A
  Magnetotransport Study}},}\ }\href@noop {} {\bibfield  {journal} {\bibinfo
  {journal} {Phys. Rev. Lett.}\ }\textbf {\bibinfo {volume} {104}},\ \bibinfo
  {pages} {126802} (\bibinfo {year} {2010})}\BibitemShut {NoStop}%
\bibitem [{\citenamefont {Liang}\ \emph {et~al.}(2015)\citenamefont {Liang},
  \citenamefont {Cheng}, \citenamefont {Wei}, \citenamefont {Luo},
  \citenamefont {Yu}, \citenamefont {Zeng},\ and\ \citenamefont
  {Zhang}}]{Liang:2015fy}%
  \BibitemOpen
  \bibfield  {author} {\bibinfo {author} {\bibfnamefont {Haixing}\ \bibnamefont
  {Liang}}, \bibinfo {author} {\bibfnamefont {Long}\ \bibnamefont {Cheng}},
  \bibinfo {author} {\bibfnamefont {Laiming}\ \bibnamefont {Wei}}, \bibinfo
  {author} {\bibfnamefont {Zhenlin}\ \bibnamefont {Luo}}, \bibinfo {author}
  {\bibfnamefont {Guolin}\ \bibnamefont {Yu}}, \bibinfo {author} {\bibfnamefont
  {Changgan}\ \bibnamefont {Zeng}}, \ and\ \bibinfo {author} {\bibfnamefont
  {Zhenyu}\ \bibnamefont {Zhang}},\ }\bibfield  {title} {\enquote {\bibinfo
  {title} {{Nonmonotonically tunable Rashba spin-orbit coupling by
  multiple-band filling control in SrTiO$_3$-based interfacial d-electron
  gases}},}\ }\href@noop {} {\bibfield  {journal} {\bibinfo  {journal} {Phys.
  Rev. B}\ }\textbf {\bibinfo {volume} {92}},\ \bibinfo {pages} {3414}
  (\bibinfo {year} {2015})}\BibitemShut {NoStop}%
\bibitem [{\citenamefont {Joshua}\ \emph {et~al.}(2013)\citenamefont {Joshua},
  \citenamefont {Ruhman}, \citenamefont {Pecker}, \citenamefont {Altman},\ and\
  \citenamefont {Ilani}}]{Joshua:2013wl}%
  \BibitemOpen
  \bibfield  {author} {\bibinfo {author} {\bibfnamefont {Arjun}\ \bibnamefont
  {Joshua}}, \bibinfo {author} {\bibfnamefont {Jonathan}\ \bibnamefont
  {Ruhman}}, \bibinfo {author} {\bibfnamefont {Sharon}\ \bibnamefont {Pecker}},
  \bibinfo {author} {\bibfnamefont {Ehud}\ \bibnamefont {Altman}}, \ and\
  \bibinfo {author} {\bibfnamefont {Shahal}\ \bibnamefont {Ilani}},\ }\bibfield
   {title} {\enquote {\bibinfo {title} {{Gate-tunable polarized phase of
  two-dimensional electrons at the LaAlO$_3$/SrTiO$_3$ interface}},}\
  }\href@noop {} {\bibfield  {journal} {\bibinfo  {journal} {Proc. Nat. Acad.
  Sci.}\ }\textbf {\bibinfo {volume} {110}},\ \bibinfo {pages} {9633--9638}
  (\bibinfo {year} {2013})}\BibitemShut {NoStop}%
\bibitem [{\citenamefont {Richter}\ \emph {et~al.}(2013)\citenamefont
  {Richter}, \citenamefont {Boschker}, \citenamefont {Dietsche}, \citenamefont
  {Fillis-Tsirakis}, \citenamefont {Jany}, \citenamefont {Loder}, \citenamefont
  {Kourkoutis}, \citenamefont {Muller}, \citenamefont {Kirtley}, \citenamefont
  {Schneider},\ and\ \citenamefont {Mannhart}}]{Richter:2013gn}%
  \BibitemOpen
  \bibfield  {author} {\bibinfo {author} {\bibfnamefont {C.}~\bibnamefont
  {Richter}}, \bibinfo {author} {\bibfnamefont {H.}~\bibnamefont {Boschker}},
  \bibinfo {author} {\bibfnamefont {W.}~\bibnamefont {Dietsche}}, \bibinfo
  {author} {\bibfnamefont {E.}~\bibnamefont {Fillis-Tsirakis}}, \bibinfo
  {author} {\bibfnamefont {R.}~\bibnamefont {Jany}}, \bibinfo {author}
  {\bibfnamefont {F.}~\bibnamefont {Loder}}, \bibinfo {author} {\bibfnamefont
  {L.~F.}\ \bibnamefont {Kourkoutis}}, \bibinfo {author} {\bibfnamefont
  {D.~A.}\ \bibnamefont {Muller}}, \bibinfo {author} {\bibfnamefont {J.~R.}\
  \bibnamefont {Kirtley}}, \bibinfo {author} {\bibfnamefont {C.~W.}\
  \bibnamefont {Schneider}}, \ and\ \bibinfo {author} {\bibfnamefont
  {J.}~\bibnamefont {Mannhart}},\ }\bibfield  {title} {\enquote {\bibinfo
  {title} {{Interface superconductor with gap behaviour like a high-temperature
  superconductor}},}\ }\href@noop {} {\bibfield  {journal} {\bibinfo  {journal}
  {Nature}\ }\textbf {\bibinfo {volume} {502}},\ \bibinfo {pages} {528--531}
  (\bibinfo {year} {2013})}\BibitemShut {NoStop}%
\bibitem [{\citenamefont {Cheng}\ \emph {et~al.}(2016)\citenamefont {Cheng},
  \citenamefont {Tomczyk}, \citenamefont {Tacla}, \citenamefont {Lee},
  \citenamefont {Lu}, \citenamefont {Veazey}, \citenamefont {Huang},
  \citenamefont {Irvin}, \citenamefont {Ryu}, \citenamefont {Eom},
  \citenamefont {Daley}, \citenamefont {Pekker},\ and\ \citenamefont
  {Levy}}]{Cheng:2016tk}%
  \BibitemOpen
  \bibfield  {author} {\bibinfo {author} {\bibfnamefont {Guanglei}\
  \bibnamefont {Cheng}}, \bibinfo {author} {\bibfnamefont {Michelle}\
  \bibnamefont {Tomczyk}}, \bibinfo {author} {\bibfnamefont {Alexandre~B}\
  \bibnamefont {Tacla}}, \bibinfo {author} {\bibfnamefont {Hyungwoo}\
  \bibnamefont {Lee}}, \bibinfo {author} {\bibfnamefont {Shicheng}\
  \bibnamefont {Lu}}, \bibinfo {author} {\bibfnamefont {Josh~.P}\ \bibnamefont
  {Veazey}}, \bibinfo {author} {\bibfnamefont {Mengchen}\ \bibnamefont
  {Huang}}, \bibinfo {author} {\bibfnamefont {Patrick}\ \bibnamefont {Irvin}},
  \bibinfo {author} {\bibfnamefont {Sangwoo}\ \bibnamefont {Ryu}}, \bibinfo
  {author} {\bibfnamefont {Chang-Beom}\ \bibnamefont {Eom}}, \bibinfo {author}
  {\bibfnamefont {Andrew}\ \bibnamefont {Daley}}, \bibinfo {author}
  {\bibfnamefont {David}\ \bibnamefont {Pekker}}, \ and\ \bibinfo {author}
  {\bibfnamefont {Jeremy}\ \bibnamefont {Levy}},\ }\bibfield  {title} {\enquote
  {\bibinfo {title} {{Tunable Electron-Electron Interactions in
  ${\mathrm{LaAlO}}_{3}/{\mathrm{SrTiO}}_{3}$ Nanostructures}},}\ }\href@noop
  {} {\bibfield  {journal} {\bibinfo  {journal} {Phys. Rev. X}\ }\textbf
  {\bibinfo {volume} {6}},\ \bibinfo {pages} {041042} (\bibinfo {year}
  {2016})}\BibitemShut {NoStop}%
\bibitem [{\citenamefont {Joshua}\ \emph {et~al.}(2012)\citenamefont {Joshua},
  \citenamefont {Pecker}, \citenamefont {Ruhman}, \citenamefont {Altman},\ and\
  \citenamefont {Ilani}}]{Joshua:2012bl}%
  \BibitemOpen
  \bibfield  {author} {\bibinfo {author} {\bibfnamefont {Arjun}\ \bibnamefont
  {Joshua}}, \bibinfo {author} {\bibfnamefont {S.}~\bibnamefont {Pecker}},
  \bibinfo {author} {\bibfnamefont {J.}~\bibnamefont {Ruhman}}, \bibinfo
  {author} {\bibfnamefont {E.}~\bibnamefont {Altman}}, \ and\ \bibinfo {author}
  {\bibfnamefont {S.}~\bibnamefont {Ilani}},\ }\bibfield  {title} {\enquote
  {\bibinfo {title} {{A universal critical density underlying the physics of
  electrons at the LaAlO$_3$/SrTiO$_3$ interface}},}\ }\href@noop {} {\bibfield
   {journal} {\bibinfo  {journal} {Nature Commun.}\ }\textbf {\bibinfo {volume}
  {3}},\ \bibinfo {pages} {1129} (\bibinfo {year} {2012})}\BibitemShut
  {NoStop}%
\bibitem [{\citenamefont {Smink}\ \emph {et~al.}(2017)\citenamefont {Smink},
  \citenamefont {de~Boer}, \citenamefont {Stehno}, \citenamefont {Brinkman},
  \citenamefont {van~der Wiel},\ and\ \citenamefont
  {Hilgenkamp}}]{Smink:2017cc}%
  \BibitemOpen
  \bibfield  {author} {\bibinfo {author} {\bibfnamefont {A.~E.~M.}\
  \bibnamefont {Smink}}, \bibinfo {author} {\bibfnamefont {J.~C.}\ \bibnamefont
  {de~Boer}}, \bibinfo {author} {\bibfnamefont {M.~P.}\ \bibnamefont {Stehno}},
  \bibinfo {author} {\bibfnamefont {A.}~\bibnamefont {Brinkman}}, \bibinfo
  {author} {\bibfnamefont {W.~G.}\ \bibnamefont {van~der Wiel}}, \ and\
  \bibinfo {author} {\bibfnamefont {H.}~\bibnamefont {Hilgenkamp}},\ }\bibfield
   {title} {\enquote {\bibinfo {title} {Gate-tunable band structure of the
  {LaAlO$_3$/SrTiO$_3$} interface},}\ }\href@noop {} {\bibfield  {journal}
  {\bibinfo  {journal} {Phys. Rev. Lett.}\ }\textbf {\bibinfo {volume} {118}},\
  \bibinfo {pages} {106401} (\bibinfo {year} {2017})}\BibitemShut {NoStop}%
\bibitem [{\citenamefont {Niu}\ \emph {et~al.}(2017)\citenamefont {Niu},
  \citenamefont {Zhang}, \citenamefont {Gan}, \citenamefont {Christensen},
  \citenamefont {Soosten}, \citenamefont {Garcia-Suarez}, \citenamefont
  {Riisager}, \citenamefont {Wang}, \citenamefont {Xu}, \citenamefont {Zhang},
  \citenamefont {Pryds},\ and\ \citenamefont {Chen}}]{Niu:2017ih}%
  \BibitemOpen
  \bibfield  {author} {\bibinfo {author} {\bibfnamefont {Wei}\ \bibnamefont
  {Niu}}, \bibinfo {author} {\bibfnamefont {Yu}~\bibnamefont {Zhang}}, \bibinfo
  {author} {\bibfnamefont {Yulin}\ \bibnamefont {Gan}}, \bibinfo {author}
  {\bibfnamefont {Dennis~V.}\ \bibnamefont {Christensen}}, \bibinfo {author}
  {\bibfnamefont {Merlin~V.}\ \bibnamefont {Soosten}}, \bibinfo {author}
  {\bibfnamefont {Eduardo~J.}\ \bibnamefont {Garcia-Suarez}}, \bibinfo {author}
  {\bibfnamefont {Anders}\ \bibnamefont {Riisager}}, \bibinfo {author}
  {\bibfnamefont {Xuefeng}\ \bibnamefont {Wang}}, \bibinfo {author}
  {\bibfnamefont {Yongbing}\ \bibnamefont {Xu}}, \bibinfo {author}
  {\bibfnamefont {Rong}\ \bibnamefont {Zhang}}, \bibinfo {author}
  {\bibfnamefont {Nini}\ \bibnamefont {Pryds}}, \ and\ \bibinfo {author}
  {\bibfnamefont {Yunzhong}\ \bibnamefont {Chen}},\ }\bibfield  {title}
  {\enquote {\bibinfo {title} {Giant tunability of the two-dimensional electron
  gas at the interface of {$\gamma$-Al$_2$O$_3$/SrTiO$_3$}},}\ }\href@noop {}
  {\bibfield  {journal} {\bibinfo  {journal} {Nano Lett.}\ }\textbf {\bibinfo
  {volume} {17}},\ \bibinfo {pages} {6878--6885} (\bibinfo {year}
  {2017})}\BibitemShut {NoStop}%
\bibitem [{\citenamefont {Smink}\ \emph {et~al.}(2018)\citenamefont {Smink},
  \citenamefont {Stehno}, \citenamefont {de~Boer}, \citenamefont {Brinkman},
  \citenamefont {van~der Wiel},\ and\ \citenamefont
  {Hilgenkamp}}]{Smink:2018tu}%
  \BibitemOpen
  \bibfield  {author} {\bibinfo {author} {\bibfnamefont {A.~E.~M.}\
  \bibnamefont {Smink}}, \bibinfo {author} {\bibfnamefont {M.~P.}\ \bibnamefont
  {Stehno}}, \bibinfo {author} {\bibfnamefont {J.~C.}\ \bibnamefont {de~Boer}},
  \bibinfo {author} {\bibfnamefont {A.}~\bibnamefont {Brinkman}}, \bibinfo
  {author} {\bibfnamefont {W.~G.}\ \bibnamefont {van~der Wiel}}, \ and\
  \bibinfo {author} {\bibfnamefont {H.}~\bibnamefont {Hilgenkamp}},\ }\bibfield
   {title} {\enquote {\bibinfo {title} {{Correlation between superconductivity,
  band filling and electron confinement at the LaAlO$_{3}$-SrTiO$_{3}$
  interface}},}\ }\href@noop {} {\bibfield  {journal} {\bibinfo  {journal}
  {Phys. Rev. B}\ }\textbf {\bibinfo {volume} {97}},\ \bibinfo {pages} {245113}
  (\bibinfo {year} {2018})}\BibitemShut {NoStop}%
\bibitem [{\citenamefont {Kim}\ \emph {et~al.}(2013)\citenamefont {Kim},
  \citenamefont {Lutchyn},\ and\ \citenamefont {Nayak}}]{Kim:2013vz}%
  \BibitemOpen
  \bibfield  {author} {\bibinfo {author} {\bibfnamefont {Younghyun}\
  \bibnamefont {Kim}}, \bibinfo {author} {\bibfnamefont {Roman~M}\ \bibnamefont
  {Lutchyn}}, \ and\ \bibinfo {author} {\bibfnamefont {Chetan}\ \bibnamefont
  {Nayak}},\ }\bibfield  {title} {\enquote {\bibinfo {title} {{Origin and
  transport signatures of spin-orbit Interactions in one- and two-dimensional
  SrTiO$_3$-based heterostructures}},}\ }\href@noop {} {\bibfield  {journal}
  {\bibinfo  {journal} {Phys. Rev. B}\ }\textbf {\bibinfo {volume} {87}},\
  \bibinfo {pages} {245121} (\bibinfo {year} {2013})}\BibitemShut {NoStop}%
\bibitem [{\citenamefont {Zhong}\ \emph {et~al.}(2013)\citenamefont {Zhong},
  \citenamefont {T{\'o}th},\ and\ \citenamefont {Held}}]{Zhong:2013fv}%
  \BibitemOpen
  \bibfield  {author} {\bibinfo {author} {\bibfnamefont {Zhicheng}\
  \bibnamefont {Zhong}}, \bibinfo {author} {\bibfnamefont {Anna}\ \bibnamefont
  {T{\'o}th}}, \ and\ \bibinfo {author} {\bibfnamefont {Karsten}\ \bibnamefont
  {Held}},\ }\bibfield  {title} {\enquote {\bibinfo {title} {{Theory of
  spin-orbit coupling at LaAlO$_{3}$/SrTiO$_{3}$ interfaces and SrTiO$_{3}$
  surfaces}},}\ }\href@noop {} {\bibfield  {journal} {\bibinfo  {journal}
  {Phys. Rev. B}\ }\textbf {\bibinfo {volume} {87}},\ \bibinfo {pages} {161102}
  (\bibinfo {year} {2013})}\BibitemShut {NoStop}%
\bibitem [{\citenamefont {Khalsa}\ \emph {et~al.}(2013)\citenamefont {Khalsa},
  \citenamefont {Lee},\ and\ \citenamefont {MacDonald}}]{Khalsa:2013hk}%
  \BibitemOpen
  \bibfield  {author} {\bibinfo {author} {\bibfnamefont {Guru}\ \bibnamefont
  {Khalsa}}, \bibinfo {author} {\bibfnamefont {Byounghak}\ \bibnamefont {Lee}},
  \ and\ \bibinfo {author} {\bibfnamefont {A.~H.}\ \bibnamefont {MacDonald}},\
  }\bibfield  {title} {\enquote {\bibinfo {title} {{Theory of $t_{2g}$
  electron-gas Rashba interactions}},}\ }\href@noop {} {\bibfield  {journal}
  {\bibinfo  {journal} {Phys. Rev. B}\ }\textbf {\bibinfo {volume} {88}},\
  \bibinfo {pages} {041302} (\bibinfo {year} {2013})}\BibitemShut {NoStop}%
\bibitem [{\citenamefont {Fischer}\ \emph {et~al.}(2013)\citenamefont
  {Fischer}, \citenamefont {Raghu},\ and\ \citenamefont
  {Kim}}]{Fischer:2013gg}%
  \BibitemOpen
  \bibfield  {author} {\bibinfo {author} {\bibfnamefont {Mark~H.}\ \bibnamefont
  {Fischer}}, \bibinfo {author} {\bibfnamefont {Srinivas}\ \bibnamefont
  {Raghu}}, \ and\ \bibinfo {author} {\bibfnamefont {Eun-Ah}\ \bibnamefont
  {Kim}},\ }\bibfield  {title} {\enquote {\bibinfo {title} {{Spin--orbit
  coupling in LaAlO$_3$/SrTiO$_3$ interfaces: magnetism and orbital
  ordering}},}\ }\href@noop {} {\bibfield  {journal} {\bibinfo  {journal} {New
  J. Phys.}\ }\textbf {\bibinfo {volume} {15}},\ \bibinfo {pages} {023022}
  (\bibinfo {year} {2013})}\BibitemShut {NoStop}%
\bibitem [{\citenamefont {Zhou}\ \emph {et~al.}(2015)\citenamefont {Zhou},
  \citenamefont {Shan},\ and\ \citenamefont {Xiao}}]{Zhou:2015uo}%
  \BibitemOpen
  \bibfield  {author} {\bibinfo {author} {\bibfnamefont {Jianhui}\ \bibnamefont
  {Zhou}}, \bibinfo {author} {\bibfnamefont {Wen-Yu}\ \bibnamefont {Shan}}, \
  and\ \bibinfo {author} {\bibfnamefont {Di}~\bibnamefont {Xiao}},\ }\bibfield
  {title} {\enquote {\bibinfo {title} {{Spin responses and effective
  Hamiltonian for the two dimensional electron gas at oxide interface
  {LaAlO}$_3$/{SrTiO}$_3$}},}\ }\href@noop {} {\bibfield  {journal} {\bibinfo
  {journal} {Phys. Rev. B}\ }\textbf {\bibinfo {volume} {91}},\ \bibinfo
  {pages} {241302} (\bibinfo {year} {2015})}\BibitemShut {NoStop}%
\bibitem [{\citenamefont {Nakamura}\ and\ \citenamefont
  {Yanase}(2013)}]{Nakamura:2013cb}%
  \BibitemOpen
  \bibfield  {author} {\bibinfo {author} {\bibfnamefont {Yasuharu}\
  \bibnamefont {Nakamura}}\ and\ \bibinfo {author} {\bibfnamefont {Youichi}\
  \bibnamefont {Yanase}},\ }\bibfield  {title} {\enquote {\bibinfo {title}
  {{Multi-Orbital Superconductivity in SrTiO$_3$/LaAlO$_3$ Interface and
  SrTiO$_3$ Surface}},}\ }\href@noop {} {\bibfield  {journal} {\bibinfo
  {journal} {J. Phys. Soc. Jpn.}\ }\textbf {\bibinfo {volume} {82}},\ \bibinfo
  {pages} {083705} (\bibinfo {year} {2013})}\BibitemShut {NoStop}%
\bibitem [{\citenamefont {Nandy}\ \emph {et~al.}(2016)\citenamefont {Nandy},
  \citenamefont {Mohanta}, \citenamefont {Acharya},\ and\ \citenamefont
  {Taraphder}}]{Nandy:2016jm}%
  \BibitemOpen
  \bibfield  {author} {\bibinfo {author} {\bibfnamefont {S.}~\bibnamefont
  {Nandy}}, \bibinfo {author} {\bibfnamefont {N.}~\bibnamefont {Mohanta}},
  \bibinfo {author} {\bibfnamefont {S.}~\bibnamefont {Acharya}}, \ and\
  \bibinfo {author} {\bibfnamefont {A.}~\bibnamefont {Taraphder}},\ }\bibfield
  {title} {\enquote {\bibinfo {title} {{Anomalous transport near the Lifshitz
  transition at the LaAlO$_3$/SrTiO$_3$ interface}},}\ }\href@noop {}
  {\bibfield  {journal} {\bibinfo  {journal} {Phys. Rev. B}\ }\textbf {\bibinfo
  {volume} {94}},\ \bibinfo {pages} {155103} (\bibinfo {year}
  {2016})}\BibitemShut {NoStop}%
\bibitem [{\citenamefont {Popovi{\'c}}\ \emph {et~al.}(2008)\citenamefont
  {Popovi{\'c}}, \citenamefont {Satpathy},\ and\ \citenamefont
  {Martin}}]{Popovic:2008ft}%
  \BibitemOpen
  \bibfield  {author} {\bibinfo {author} {\bibfnamefont {Zoran}\ \bibnamefont
  {Popovi{\'c}}}, \bibinfo {author} {\bibfnamefont {Sashi}\ \bibnamefont
  {Satpathy}}, \ and\ \bibinfo {author} {\bibfnamefont {Richard}\ \bibnamefont
  {Martin}},\ }\bibfield  {title} {\enquote {\bibinfo {title} {Origin of the
  two-dimensional electron gas carrier density at the {LaAlO$_3$ on SrTiO$_3$}
  interface},}\ }\href@noop {} {\bibfield  {journal} {\bibinfo  {journal}
  {Phys. Rev. Lett.}\ }\textbf {\bibinfo {volume} {101}},\ \bibinfo {pages}
  {256801} (\bibinfo {year} {2008})}\BibitemShut {NoStop}%
\bibitem [{\citenamefont {Son}\ \emph {et~al.}(2009)\citenamefont {Son},
  \citenamefont {Cho}, \citenamefont {Lee}, \citenamefont {Lee},\ and\
  \citenamefont {Han}}]{Son:2009wb}%
  \BibitemOpen
  \bibfield  {author} {\bibinfo {author} {\bibfnamefont {Won-Joon}\
  \bibnamefont {Son}}, \bibinfo {author} {\bibfnamefont {Eunae}\ \bibnamefont
  {Cho}}, \bibinfo {author} {\bibfnamefont {Bora}\ \bibnamefont {Lee}},
  \bibinfo {author} {\bibfnamefont {Jaichan}\ \bibnamefont {Lee}}, \ and\
  \bibinfo {author} {\bibfnamefont {Seungwu}\ \bibnamefont {Han}},\ }\bibfield
  {title} {\enquote {\bibinfo {title} {{Density and spatial distribution of
  charge carriers in the intrinsic \emph{n}-type LaAlO$_{3}$-SrTiO$_{3}$
  interface}},}\ }\href@noop {} {\bibfield  {journal} {\bibinfo  {journal}
  {Phys. Rev. B}\ }\textbf {\bibinfo {volume} {79}},\ \bibinfo {pages} {245411}
  (\bibinfo {year} {2009})}\BibitemShut {NoStop}%
\bibitem [{\citenamefont {Pentcheva}\ and\ \citenamefont
  {Pickett}(2009)}]{Pentcheva:2009ef}%
  \BibitemOpen
  \bibfield  {author} {\bibinfo {author} {\bibfnamefont {Rossitza}\
  \bibnamefont {Pentcheva}}\ and\ \bibinfo {author} {\bibfnamefont {Warren~E.}\
  \bibnamefont {Pickett}},\ }\bibfield  {title} {\enquote {\bibinfo {title}
  {Avoiding the polarization catastrophe in {LaAlO$_3$} overlayers on
  {SrTiO$_3$}(001) through polar distortion},}\ }\href@noop {} {\bibfield
  {journal} {\bibinfo  {journal} {Phys. Rev. Lett.}\ }\textbf {\bibinfo
  {volume} {102}},\ \bibinfo {pages} {107602} (\bibinfo {year}
  {2009})}\BibitemShut {NoStop}%
\bibitem [{\citenamefont {Stengel}(2011)}]{Stengel:2011hy}%
  \BibitemOpen
  \bibfield  {author} {\bibinfo {author} {\bibfnamefont {Massimiliano}\
  \bibnamefont {Stengel}},\ }\bibfield  {title} {\enquote {\bibinfo {title}
  {First-principles modeling of electrostatically doped perovskite systems},}\
  }\href@noop {} {\bibfield  {journal} {\bibinfo  {journal} {Phys. Rev. Lett.}\
  }\textbf {\bibinfo {volume} {106}},\ \bibinfo {pages} {136803} (\bibinfo
  {year} {2011})}\BibitemShut {NoStop}%
\bibitem [{\citenamefont {Khalsa}\ and\ \citenamefont
  {MacDonald}(2012)}]{Khalsa:2012fu}%
  \BibitemOpen
  \bibfield  {author} {\bibinfo {author} {\bibfnamefont {Guru}\ \bibnamefont
  {Khalsa}}\ and\ \bibinfo {author} {\bibfnamefont {A.~H.}\ \bibnamefont
  {MacDonald}},\ }\bibfield  {title} {\enquote {\bibinfo {title} {{Theory of
  the SrTiO$_{3}$ surface state two-dimensional electron gas}},}\ }\href@noop
  {} {\bibfield  {journal} {\bibinfo  {journal} {Phys. Rev. B}\ }\textbf
  {\bibinfo {volume} {86}},\ \bibinfo {pages} {125121} (\bibinfo {year}
  {2012})}\BibitemShut {NoStop}%
\bibitem [{\citenamefont {Raslan}\ \emph {et~al.}(2017)\citenamefont {Raslan},
  \citenamefont {Lafleur},\ and\ \citenamefont {Atkinson}}]{Raslan:2017gh}%
  \BibitemOpen
  \bibfield  {author} {\bibinfo {author} {\bibfnamefont {Amany}\ \bibnamefont
  {Raslan}}, \bibinfo {author} {\bibfnamefont {Patrick}\ \bibnamefont
  {Lafleur}}, \ and\ \bibinfo {author} {\bibfnamefont {W.~A.}\ \bibnamefont
  {Atkinson}},\ }\bibfield  {title} {\enquote {\bibinfo {title}
  {{Temperature-dependent band structure of SrTiO$_3$ interfaces}},}\
  }\href@noop {} {\bibfield  {journal} {\bibinfo  {journal} {Phys. Rev. B}\
  }\textbf {\bibinfo {volume} {95}},\ \bibinfo {pages} {054106} (\bibinfo
  {year} {2017})}\BibitemShut {NoStop}%
\bibitem [{\citenamefont {Li}\ \emph {et~al.}(2018)\citenamefont {Li},
  \citenamefont {Lemal}, \citenamefont {Gariglio}, \citenamefont {Wu},
  \citenamefont {F{\^e}te}, \citenamefont {Boselli}, \citenamefont {Ghosez},\
  and\ \citenamefont {Triscone}}]{Li:2018hy}%
  \BibitemOpen
  \bibfield  {author} {\bibinfo {author} {\bibfnamefont {Danfeng}\ \bibnamefont
  {Li}}, \bibinfo {author} {\bibfnamefont {S{\'e}bastien}\ \bibnamefont
  {Lemal}}, \bibinfo {author} {\bibfnamefont {Stefano}\ \bibnamefont
  {Gariglio}}, \bibinfo {author} {\bibfnamefont {Zhenping}\ \bibnamefont {Wu}},
  \bibinfo {author} {\bibfnamefont {Alexandre}\ \bibnamefont {F{\^e}te}},
  \bibinfo {author} {\bibfnamefont {Margherita}\ \bibnamefont {Boselli}},
  \bibinfo {author} {\bibfnamefont {Philippe}\ \bibnamefont {Ghosez}}, \ and\
  \bibinfo {author} {\bibfnamefont {Jean-Marc}\ \bibnamefont {Triscone}},\
  }\bibfield  {title} {\enquote {\bibinfo {title} {{Probing Quantum Confinement
  and Electronic Structure at Polar Oxide Interfaces}},}\ }\href@noop {}
  {\bibfield  {journal} {\bibinfo  {journal} {Advanced Science}\ }\textbf
  {\bibinfo {volume} {4}},\ \bibinfo {pages} {1800242} (\bibinfo {year}
  {2018})}\BibitemShut {NoStop}%
\bibitem [{\citenamefont {Rowley}\ \emph {et~al.}(2014)\citenamefont {Rowley},
  \citenamefont {Spalek}, \citenamefont {Smith}, \citenamefont {Dean},
  \citenamefont {Itoh}, \citenamefont {Scott}, \citenamefont {Lonzarich},\ and\
  \citenamefont {Saxena}}]{Rowley:2014bda}%
  \BibitemOpen
  \bibfield  {author} {\bibinfo {author} {\bibfnamefont {S.~E.}\ \bibnamefont
  {Rowley}}, \bibinfo {author} {\bibfnamefont {L.~J.}\ \bibnamefont {Spalek}},
  \bibinfo {author} {\bibfnamefont {R.~P.}\ \bibnamefont {Smith}}, \bibinfo
  {author} {\bibfnamefont {M.~P.~M.}\ \bibnamefont {Dean}}, \bibinfo {author}
  {\bibfnamefont {M.}~\bibnamefont {Itoh}}, \bibinfo {author} {\bibfnamefont
  {J.~F.}\ \bibnamefont {Scott}}, \bibinfo {author} {\bibfnamefont {G.~G.}\
  \bibnamefont {Lonzarich}}, \ and\ \bibinfo {author} {\bibfnamefont {S.~S.}\
  \bibnamefont {Saxena}},\ }\bibfield  {title} {\enquote {\bibinfo {title}
  {{Ferroelectric quantum criticality}},}\ }\href@noop {} {\bibfield  {journal}
  {\bibinfo  {journal} {Nature Phys.}\ }\textbf {\bibinfo {volume} {10}},\
  \bibinfo {pages} {367--372} (\bibinfo {year} {2014})}\BibitemShut {NoStop}%
\bibitem [{\citenamefont {Atkinson}\ \emph {et~al.}(2017)\citenamefont
  {Atkinson}, \citenamefont {Lafleur},\ and\ \citenamefont
  {Raslan}}]{Atkinson:2017jt}%
  \BibitemOpen
  \bibfield  {author} {\bibinfo {author} {\bibfnamefont {W.~A.}\ \bibnamefont
  {Atkinson}}, \bibinfo {author} {\bibfnamefont {Patrick}\ \bibnamefont
  {Lafleur}}, \ and\ \bibinfo {author} {\bibfnamefont {Amany}\ \bibnamefont
  {Raslan}},\ }\bibfield  {title} {\enquote {\bibinfo {title} {{Influence of
  the ferroelectric quantum critical point on SrTiO$_3$ interfaces}},}\
  }\href@noop {} {\bibfield  {journal} {\bibinfo  {journal} {Phys. Rev. B}\
  }\textbf {\bibinfo {volume} {95}},\ \bibinfo {pages} {054107} (\bibinfo
  {year} {2017})}\BibitemShut {NoStop}%
\bibitem [{\citenamefont {Zubko}\ \emph {et~al.}(2013)\citenamefont {Zubko},
  \citenamefont {Catalan},\ and\ \citenamefont {Tagantsev}}]{Zubko:2013bt}%
  \BibitemOpen
  \bibfield  {author} {\bibinfo {author} {\bibfnamefont {Pavlo}\ \bibnamefont
  {Zubko}}, \bibinfo {author} {\bibfnamefont {Gustau}\ \bibnamefont {Catalan}},
  \ and\ \bibinfo {author} {\bibfnamefont {Alexander~K}\ \bibnamefont
  {Tagantsev}},\ }\bibfield  {title} {\enquote {\bibinfo {title}
  {{Flexoelectric Effect in Solids}},}\ }\href@noop {} {\bibfield  {journal}
  {\bibinfo  {journal} {Ann. Rev. Mat. Res.}\ }\textbf {\bibinfo {volume}
  {43}},\ \bibinfo {pages} {387--421} (\bibinfo {year} {2013})}\BibitemShut
  {NoStop}%
\bibitem [{\citenamefont {Uwe}\ and\ \citenamefont
  {Sakudo}(1976)}]{Uwe:1976fj}%
  \BibitemOpen
  \bibfield  {author} {\bibinfo {author} {\bibfnamefont {Hiromoto}\
  \bibnamefont {Uwe}}\ and\ \bibinfo {author} {\bibfnamefont {Tunetaro}\
  \bibnamefont {Sakudo}},\ }\bibfield  {title} {\enquote {\bibinfo {title}
  {{Stress-induced ferroelectricity and soft phonon modes in SrTiO$_3$}},}\
  }\href@noop {} {\bibfield  {journal} {\bibinfo  {journal} {Phys. Rev. B}\
  }\textbf {\bibinfo {volume} {13}},\ \bibinfo {pages} {271--286} (\bibinfo
  {year} {1976})}\BibitemShut {NoStop}%
\bibitem [{\citenamefont {Wang}\ \emph {et~al.}(2000)\citenamefont {Wang},
  \citenamefont {Sakamoto},\ and\ \citenamefont {Itoh}}]{Wang:2000vy}%
  \BibitemOpen
  \bibfield  {author} {\bibinfo {author} {\bibfnamefont {Ruiping}\ \bibnamefont
  {Wang}}, \bibinfo {author} {\bibfnamefont {Norihiko}\ \bibnamefont
  {Sakamoto}}, \ and\ \bibinfo {author} {\bibfnamefont {Mitsuru}\ \bibnamefont
  {Itoh}},\ }\bibfield  {title} {\enquote {\bibinfo {title} {{Effects of
  pressure on the dielectric properties of SrTi${}^{18}$O$_{3 }$ and
  SrTi${}^{16}$O$_{3 }$ single crystals }},}\ }\href@noop {} {\bibfield
  {journal} {\bibinfo  {journal} {Phys. Rev. B}\ }\textbf {\bibinfo {volume}
  {62}},\ \bibinfo {pages} {3577--3580} (\bibinfo {year} {2000})}\BibitemShut
  {NoStop}%
\bibitem [{\citenamefont {Haeni}\ \emph {et~al.}(2004)\citenamefont {Haeni},
  \citenamefont {Irvin}, \citenamefont {Chang}, \citenamefont {Uecker},
  \citenamefont {Reiche}, \citenamefont {Li}, \citenamefont {Choudhury},
  \citenamefont {Tian}, \citenamefont {Hawley}, \citenamefont {Craigo},
  \citenamefont {Tagantsev}, \citenamefont {Pan}, \citenamefont {Streiffer},
  \citenamefont {Chen}, \citenamefont {Kirchoefer}, \citenamefont {Levy},\ and\
  \citenamefont {Schlom}}]{Haeni:2004gj}%
  \BibitemOpen
  \bibfield  {author} {\bibinfo {author} {\bibfnamefont {J.~H.}\ \bibnamefont
  {Haeni}}, \bibinfo {author} {\bibfnamefont {P.}~\bibnamefont {Irvin}},
  \bibinfo {author} {\bibfnamefont {W.}~\bibnamefont {Chang}}, \bibinfo
  {author} {\bibfnamefont {R.}~\bibnamefont {Uecker}}, \bibinfo {author}
  {\bibfnamefont {P.}~\bibnamefont {Reiche}}, \bibinfo {author} {\bibfnamefont
  {Y.~L.}\ \bibnamefont {Li}}, \bibinfo {author} {\bibfnamefont
  {S.}~\bibnamefont {Choudhury}}, \bibinfo {author} {\bibfnamefont
  {W.}~\bibnamefont {Tian}}, \bibinfo {author} {\bibfnamefont {M.~E.}\
  \bibnamefont {Hawley}}, \bibinfo {author} {\bibfnamefont {B.}~\bibnamefont
  {Craigo}}, \bibinfo {author} {\bibfnamefont {A.~K.}\ \bibnamefont
  {Tagantsev}}, \bibinfo {author} {\bibfnamefont {X.~Q.}\ \bibnamefont {Pan}},
  \bibinfo {author} {\bibfnamefont {S.~K.}\ \bibnamefont {Streiffer}}, \bibinfo
  {author} {\bibfnamefont {L.~Q.}\ \bibnamefont {Chen}}, \bibinfo {author}
  {\bibfnamefont {S.~W.}\ \bibnamefont {Kirchoefer}}, \bibinfo {author}
  {\bibfnamefont {J.}~\bibnamefont {Levy}}, \ and\ \bibinfo {author}
  {\bibfnamefont {D.~G.}\ \bibnamefont {Schlom}},\ }\bibfield  {title}
  {\enquote {\bibinfo {title} {{Room-temperature ferroelectricity in strained
  SrTiO$_3$}},}\ }\href@noop {} {\bibfield  {journal} {\bibinfo  {journal}
  {Nature Commun.}\ }\textbf {\bibinfo {volume} {430}},\ \bibinfo {pages}
  {758--761} (\bibinfo {year} {2004})}\BibitemShut {NoStop}%
\bibitem [{\citenamefont {Bark}\ \emph {et~al.}(2011)\citenamefont {Bark},
  \citenamefont {Felker}, \citenamefont {Wang}, \citenamefont {Zhang},
  \citenamefont {Jang}, \citenamefont {Folkman}, \citenamefont {Park},
  \citenamefont {Baek}, \citenamefont {Zhou}, \citenamefont {Fong},
  \citenamefont {Pan}, \citenamefont {Tsymbal}, \citenamefont {Rzchowski},\
  and\ \citenamefont {Eom}}]{Bark:2011fo}%
  \BibitemOpen
  \bibfield  {author} {\bibinfo {author} {\bibfnamefont {C.~W.}\ \bibnamefont
  {Bark}}, \bibinfo {author} {\bibfnamefont {D.~A.}\ \bibnamefont {Felker}},
  \bibinfo {author} {\bibfnamefont {Yuxuan}\ \bibnamefont {Wang}}, \bibinfo
  {author} {\bibfnamefont {Y.}~\bibnamefont {Zhang}}, \bibinfo {author}
  {\bibfnamefont {H.~W.}\ \bibnamefont {Jang}}, \bibinfo {author}
  {\bibfnamefont {C.~M.}\ \bibnamefont {Folkman}}, \bibinfo {author}
  {\bibfnamefont {J.~W.}\ \bibnamefont {Park}}, \bibinfo {author}
  {\bibfnamefont {S.~H.}\ \bibnamefont {Baek}}, \bibinfo {author}
  {\bibfnamefont {H.}~\bibnamefont {Zhou}}, \bibinfo {author} {\bibfnamefont
  {D.~D.}\ \bibnamefont {Fong}}, \bibinfo {author} {\bibfnamefont {X.~Q.}\
  \bibnamefont {Pan}}, \bibinfo {author} {\bibfnamefont {E.~Y.}\ \bibnamefont
  {Tsymbal}}, \bibinfo {author} {\bibfnamefont {M.~S.}\ \bibnamefont
  {Rzchowski}}, \ and\ \bibinfo {author} {\bibfnamefont {C.~B.}\ \bibnamefont
  {Eom}},\ }\bibfield  {title} {\enquote {\bibinfo {title} {{Tailoring a
  two-dimensional electron gas at the LaAlO$_3$/SrTiO$_3$ (001) interface by
  epitaxial strain}},}\ }\href@noop {} {\bibfield  {journal} {\bibinfo
  {journal} {Proc. Nat. Acad. Sci.}\ }\textbf {\bibinfo {volume} {108}},\
  \bibinfo {pages} {4720--4724} (\bibinfo {year} {2011})}\BibitemShut {NoStop}%
\bibitem [{\citenamefont {Behtash}\ \emph {et~al.}(2016)\citenamefont
  {Behtash}, \citenamefont {Nazir}, \citenamefont {Wang},\ and\ \citenamefont
  {Yang}}]{Behtash:2016dt}%
  \BibitemOpen
  \bibfield  {author} {\bibinfo {author} {\bibfnamefont {Maziar}\ \bibnamefont
  {Behtash}}, \bibinfo {author} {\bibfnamefont {Safdar}\ \bibnamefont {Nazir}},
  \bibinfo {author} {\bibfnamefont {Yaqin}\ \bibnamefont {Wang}}, \ and\
  \bibinfo {author} {\bibfnamefont {Kesong}\ \bibnamefont {Yang}},\ }\bibfield
  {title} {\enquote {\bibinfo {title} {{Polarization effects on the interfacial
  conductivity in LaAlO$_3$/SrTiO$_3$ heterostructures: a first-principles
  study}},}\ }\href@noop {} {\bibfield  {journal} {\bibinfo  {journal} {Phys.
  Chem. Chem. Phys.}\ }\textbf {\bibinfo {volume} {18}},\ \bibinfo {pages}
  {6831--6838} (\bibinfo {year} {2016})}\BibitemShut {NoStop}%
\bibitem [{\citenamefont {Seiler}\ \emph {et~al.}(2018)\citenamefont {Seiler},
  \citenamefont {Zabaleta}, \citenamefont {Wanke}, \citenamefont {Mannhart},
  \citenamefont {Kopp},\ and\ \citenamefont {Braak}}]{Seiler:2018eo}%
  \BibitemOpen
  \bibfield  {author} {\bibinfo {author} {\bibfnamefont {Patrick}\ \bibnamefont
  {Seiler}}, \bibinfo {author} {\bibfnamefont {Jone}\ \bibnamefont {Zabaleta}},
  \bibinfo {author} {\bibfnamefont {Robin}\ \bibnamefont {Wanke}}, \bibinfo
  {author} {\bibfnamefont {Jochen}\ \bibnamefont {Mannhart}}, \bibinfo {author}
  {\bibfnamefont {Thilo}\ \bibnamefont {Kopp}}, \ and\ \bibinfo {author}
  {\bibfnamefont {Daniel}\ \bibnamefont {Braak}},\ }\bibfield  {title}
  {\enquote {\bibinfo {title} {{Antilocalization at an oxide interface}},}\
  }\href@noop {} {\bibfield  {journal} {\bibinfo  {journal} {Phys. Rev. B}\
  }\textbf {\bibinfo {volume} {97}},\ \bibinfo {pages} {199} (\bibinfo {year}
  {2018})}\BibitemShut {NoStop}%
\bibitem [{\citenamefont {Willmott}\ \emph {et~al.}(2007)\citenamefont
  {Willmott}, \citenamefont {Pauli}, \citenamefont {Herger}, \citenamefont
  {Schlep{\"u}tz}, \citenamefont {Martoccia}, \citenamefont {Patterson},
  \citenamefont {Delley}, \citenamefont {Clarke}, \citenamefont {Kumah},
  \citenamefont {Cionca},\ and\ \citenamefont {Yacoby}}]{Willmott:2007ip}%
  \BibitemOpen
  \bibfield  {author} {\bibinfo {author} {\bibfnamefont {P.~R.}\ \bibnamefont
  {Willmott}}, \bibinfo {author} {\bibfnamefont {S.~A.}\ \bibnamefont {Pauli}},
  \bibinfo {author} {\bibfnamefont {R.}~\bibnamefont {Herger}}, \bibinfo
  {author} {\bibfnamefont {C.~M.}\ \bibnamefont {Schlep{\"u}tz}}, \bibinfo
  {author} {\bibfnamefont {D.}~\bibnamefont {Martoccia}}, \bibinfo {author}
  {\bibfnamefont {B.~D.}\ \bibnamefont {Patterson}}, \bibinfo {author}
  {\bibfnamefont {B.}~\bibnamefont {Delley}}, \bibinfo {author} {\bibfnamefont
  {R.}~\bibnamefont {Clarke}}, \bibinfo {author} {\bibfnamefont
  {D.}~\bibnamefont {Kumah}}, \bibinfo {author} {\bibfnamefont
  {C.}~\bibnamefont {Cionca}}, \ and\ \bibinfo {author} {\bibfnamefont
  {Y.}~\bibnamefont {Yacoby}},\ }\bibfield  {title} {\enquote {\bibinfo {title}
  {Structural basis for the conducting interface between {LaAlO$_3$and
  SrTiO$_3$}},}\ }\href@noop {} {\bibfield  {journal} {\bibinfo  {journal}
  {Phys. Rev. Lett.}\ }\textbf {\bibinfo {volume} {99}},\ \bibinfo {pages}
  {155502} (\bibinfo {year} {2007})}\BibitemShut {NoStop}%
\bibitem [{\citenamefont {Lee}\ \emph {et~al.}(2016)\citenamefont {Lee},
  \citenamefont {Singh}, \citenamefont {Liu}, \citenamefont {Lin},
  \citenamefont {Chen}, \citenamefont {Guo}, \citenamefont {Chu},\ and\
  \citenamefont {Chu}}]{Lee:2016dj}%
  \BibitemOpen
  \bibfield  {author} {\bibinfo {author} {\bibfnamefont {P.~W.}\ \bibnamefont
  {Lee}}, \bibinfo {author} {\bibfnamefont {V.~N.}\ \bibnamefont {Singh}},
  \bibinfo {author} {\bibfnamefont {H.~J.}\ \bibnamefont {Liu}}, \bibinfo
  {author} {\bibfnamefont {J.~C.}\ \bibnamefont {Lin}}, \bibinfo {author}
  {\bibfnamefont {C.~H.}\ \bibnamefont {Chen}}, \bibinfo {author}
  {\bibfnamefont {G.~Y.}\ \bibnamefont {Guo}}, \bibinfo {author} {\bibfnamefont
  {Y.~H.}\ \bibnamefont {Chu}}, \ and\ \bibinfo {author} {\bibfnamefont
  {M.~W.}\ \bibnamefont {Chu}},\ }\bibfield  {title} {\enquote {\bibinfo
  {title} {{Hidden lattice instabilities as origin of the conductive interface
  between insulating LaAlO}},}\ }\href@noop {} {\bibfield  {journal} {\bibinfo
  {journal} {Nature Commun.}\ }\textbf {\bibinfo {volume} {7}},\ \bibinfo
  {pages} {12773} (\bibinfo {year} {2016})}\BibitemShut {NoStop}%
\bibitem [{\citenamefont {Yudin}\ and\ \citenamefont
  {Tagantsev}(2013)}]{Yudin:2013}%
  \BibitemOpen
  \bibfield  {author} {\bibinfo {author} {\bibfnamefont {P.~V.}\ \bibnamefont
  {Yudin}}\ and\ \bibinfo {author} {\bibfnamefont {A.~K.}\ \bibnamefont
  {Tagantsev}},\ }\bibfield  {title} {\enquote {\bibinfo {title} {{Fundamentals
  of flexoelectricity in solids}},}\ }\href@noop {} {\bibfield  {journal}
  {\bibinfo  {journal} {Nanotechnology}\ }\textbf {\bibinfo {volume} {24}},\
  \bibinfo {pages} {432001} (\bibinfo {year} {2013})}\BibitemShut {NoStop}%
\bibitem [{\citenamefont {Gu}\ \emph {et~al.}(2014)\citenamefont {Gu},
  \citenamefont {Li}, \citenamefont {Morozovska}, \citenamefont {Wang},
  \citenamefont {Eliseev}, \citenamefont {Gopalan},\ and\ \citenamefont
  {Chen}}]{Gu:2014iq}%
  \BibitemOpen
  \bibfield  {author} {\bibinfo {author} {\bibfnamefont {Yijia}\ \bibnamefont
  {Gu}}, \bibinfo {author} {\bibfnamefont {Menglei}\ \bibnamefont {Li}},
  \bibinfo {author} {\bibfnamefont {Anna~N.}\ \bibnamefont {Morozovska}},
  \bibinfo {author} {\bibfnamefont {Yi}~\bibnamefont {Wang}}, \bibinfo {author}
  {\bibfnamefont {Eugene~A.}\ \bibnamefont {Eliseev}}, \bibinfo {author}
  {\bibfnamefont {Venkatraman}\ \bibnamefont {Gopalan}}, \ and\ \bibinfo
  {author} {\bibfnamefont {Long-Qing}\ \bibnamefont {Chen}},\ }\bibfield
  {title} {\enquote {\bibinfo {title} {{Flexoelectricity and ferroelectric
  domain wall structures: Phase-field modeling and DFT calculations}},}\
  }\href@noop {} {\bibfield  {journal} {\bibinfo  {journal} {Phys. Rev. B}\
  }\textbf {\bibinfo {volume} {89}},\ \bibinfo {pages} {2069} (\bibinfo {year}
  {2014})}\BibitemShut {NoStop}%
\bibitem [{\citenamefont {Gao}\ \emph {et~al.}(2018)\citenamefont {Gao},
  \citenamefont {Yang}, \citenamefont {Ishikawa}, \citenamefont {Li},
  \citenamefont {Feng}, \citenamefont {Kumamoto}, \citenamefont {Shibata},
  \citenamefont {Yu},\ and\ \citenamefont {Ikuhara}}]{Gao:2018kk}%
  \BibitemOpen
  \bibfield  {author} {\bibinfo {author} {\bibfnamefont {Peng}\ \bibnamefont
  {Gao}}, \bibinfo {author} {\bibfnamefont {Shuzhen}\ \bibnamefont {Yang}},
  \bibinfo {author} {\bibfnamefont {Ryo}\ \bibnamefont {Ishikawa}}, \bibinfo
  {author} {\bibfnamefont {Ning}\ \bibnamefont {Li}}, \bibinfo {author}
  {\bibfnamefont {Bin}\ \bibnamefont {Feng}}, \bibinfo {author} {\bibfnamefont
  {Akihito}\ \bibnamefont {Kumamoto}}, \bibinfo {author} {\bibfnamefont
  {Naoya}\ \bibnamefont {Shibata}}, \bibinfo {author} {\bibfnamefont
  {Pu}~\bibnamefont {Yu}}, \ and\ \bibinfo {author} {\bibfnamefont {Yuichi}\
  \bibnamefont {Ikuhara}},\ }\bibfield  {title} {\enquote {\bibinfo {title}
  {{Atomic-Scale Measurement of Flexoelectric Polarization at SrTiO$_3$
  Dislocations}},}\ }\href@noop {} {\bibfield  {journal} {\bibinfo  {journal}
  {Phys. Rev. Lett.}\ }\textbf {\bibinfo {volume} {120}},\ \bibinfo {pages}
  {267601} (\bibinfo {year} {2018})}\BibitemShut {NoStop}%
\bibitem [{\citenamefont {Abdollahi}\ \emph {et~al.}(2015)\citenamefont
  {Abdollahi}, \citenamefont {Peco}, \citenamefont {Mill{\'a}n}, \citenamefont
  {Arroyo}, \citenamefont {Catalan},\ and\ \citenamefont
  {Arias}}]{Abdollahi:2015bn}%
  \BibitemOpen
  \bibfield  {author} {\bibinfo {author} {\bibfnamefont {Amir}\ \bibnamefont
  {Abdollahi}}, \bibinfo {author} {\bibfnamefont {Christian}\ \bibnamefont
  {Peco}}, \bibinfo {author} {\bibfnamefont {Daniel}\ \bibnamefont
  {Mill{\'a}n}}, \bibinfo {author} {\bibfnamefont {Marino}\ \bibnamefont
  {Arroyo}}, \bibinfo {author} {\bibfnamefont {Gustau}\ \bibnamefont
  {Catalan}}, \ and\ \bibinfo {author} {\bibfnamefont {Irene}\ \bibnamefont
  {Arias}},\ }\bibfield  {title} {\enquote {\bibinfo {title} {{Fracture
  toughening and toughness asymmetry induced by flexoelectricity}},}\
  }\href@noop {} {\bibfield  {journal} {\bibinfo  {journal} {Phys. Rev. B}\
  }\textbf {\bibinfo {volume} {92}},\ \bibinfo {pages} {2069} (\bibinfo {year}
  {2015})}\BibitemShut {NoStop}%
\bibitem [{\citenamefont {Majdoub}\ \emph {et~al.}(2009)\citenamefont
  {Majdoub}, \citenamefont {Maranganti},\ and\ \citenamefont
  {Sharma}}]{Majdoub:2009il}%
  \BibitemOpen
  \bibfield  {author} {\bibinfo {author} {\bibfnamefont {M.~S.}\ \bibnamefont
  {Majdoub}}, \bibinfo {author} {\bibfnamefont {R.}~\bibnamefont {Maranganti}},
  \ and\ \bibinfo {author} {\bibfnamefont {P.}~\bibnamefont {Sharma}},\
  }\bibfield  {title} {\enquote {\bibinfo {title} {{Understanding the origins
  of the intrinsic dead layer effect in nanocapacitors}},}\ }\href@noop {}
  {\bibfield  {journal} {\bibinfo  {journal} {Phys. Rev. B}\ }\textbf {\bibinfo
  {volume} {79}},\ \bibinfo {pages} {502} (\bibinfo {year} {2009})}\BibitemShut
  {NoStop}%
\bibitem [{\citenamefont {Nakagawa}\ \emph {et~al.}(2006)\citenamefont
  {Nakagawa}, \citenamefont {Hwang},\ and\ \citenamefont
  {Muller}}]{Nakagawa:2006gt}%
  \BibitemOpen
  \bibfield  {author} {\bibinfo {author} {\bibfnamefont {Naoyuki}\ \bibnamefont
  {Nakagawa}}, \bibinfo {author} {\bibfnamefont {Harold~Y.}\ \bibnamefont
  {Hwang}}, \ and\ \bibinfo {author} {\bibfnamefont {David~A.}\ \bibnamefont
  {Muller}},\ }\bibfield  {title} {\enquote {\bibinfo {title} {{Why some
  interfaces cannot be sharp}},}\ }\href@noop {} {\bibfield  {journal}
  {\bibinfo  {journal} {Nature Mat.}\ }\textbf {\bibinfo {volume} {5}},\
  \bibinfo {pages} {204--209} (\bibinfo {year} {2006})}\BibitemShut {NoStop}%
\bibitem [{\citenamefont {Bristowe}\ \emph {et~al.}(2014)\citenamefont
  {Bristowe}, \citenamefont {Ghosez}, \citenamefont {Littlewood},\ and\
  \citenamefont {Artacho}}]{Bristowe:2014fc}%
  \BibitemOpen
  \bibfield  {author} {\bibinfo {author} {\bibfnamefont {N.~C.}\ \bibnamefont
  {Bristowe}}, \bibinfo {author} {\bibfnamefont {Philippe}\ \bibnamefont
  {Ghosez}}, \bibinfo {author} {\bibfnamefont {P.~B.}\ \bibnamefont
  {Littlewood}}, \ and\ \bibinfo {author} {\bibfnamefont {Emilio}\ \bibnamefont
  {Artacho}},\ }\bibfield  {title} {\enquote {\bibinfo {title} {{The origin of
  two-dimensional electron gases at oxide interfaces: insights from theory}},}\
  }\href@noop {} {\bibfield  {journal} {\bibinfo  {journal} {J. Phys. Cond.
  Mat.}\ }\textbf {\bibinfo {volume} {26}},\ \bibinfo {pages} {143201}
  (\bibinfo {year} {2014})}\BibitemShut {NoStop}%
\bibitem [{\citenamefont {Piyanzina}\ \emph {et~al.}(2018)\citenamefont
  {Piyanzina}, \citenamefont {Eyert}, \citenamefont {Lysogorskiy},
  \citenamefont {Tayurskii},\ and\ \citenamefont {Kopp}}]{Piyanzina:2018ua}%
  \BibitemOpen
  \bibfield  {author} {\bibinfo {author} {\bibfnamefont {I.~I.}\ \bibnamefont
  {Piyanzina}}, \bibinfo {author} {\bibfnamefont {V.}~\bibnamefont {Eyert}},
  \bibinfo {author} {\bibfnamefont {Yu~V.}\ \bibnamefont {Lysogorskiy}},
  \bibinfo {author} {\bibfnamefont {D.~A.}\ \bibnamefont {Tayurskii}}, \ and\
  \bibinfo {author} {\bibfnamefont {T.}~\bibnamefont {Kopp}},\ }\bibfield
  {title} {\enquote {\bibinfo {title} {{Oxygen vacancies and hydrogen doping in
  LaAlO$_3$/SrTiO$_3$ heterostructures: electronic properties and impact on
  surface and interface reconstruction}},}\ }\href@noop {} {\  (\bibinfo {year}
  {2018})},\ \Eprint {http://arxiv.org/abs/arXiv:1803.01382} {arXiv:1803.01382}
  \BibitemShut {NoStop}%
\bibitem [{\citenamefont {Gariglio}\ \emph {et~al.}(2015)\citenamefont
  {Gariglio}, \citenamefont {F{\^e}te},\ and\ \citenamefont
  {Triscone}}]{Gariglio:2015jx}%
  \BibitemOpen
  \bibfield  {author} {\bibinfo {author} {\bibfnamefont {S.}~\bibnamefont
  {Gariglio}}, \bibinfo {author} {\bibfnamefont {A.}~\bibnamefont {F{\^e}te}},
  \ and\ \bibinfo {author} {\bibfnamefont {J.~M.}\ \bibnamefont {Triscone}},\
  }\bibfield  {title} {\enquote {\bibinfo {title} {{Electron confinement at the
  LaAlO$_3$/SrTiO$_3$ interface}},}\ }\href@noop {} {\bibfield  {journal}
  {\bibinfo  {journal} {J. Phys. Cond. Mat.}\ }\textbf {\bibinfo {volume}
  {27}},\ \bibinfo {pages} {283201} (\bibinfo {year} {2015})}\BibitemShut
  {NoStop}%
\bibitem [{\citenamefont {Allen}\ \emph {et~al.}(2013)\citenamefont {Allen},
  \citenamefont {Jalan}, \citenamefont {Lee}, \citenamefont {Ouellette},
  \citenamefont {Khalsa}, \citenamefont {Jaroszynski}, \citenamefont
  {Stemmer},\ and\ \citenamefont {MacDonald}}]{Allen:2013wk}%
  \BibitemOpen
  \bibfield  {author} {\bibinfo {author} {\bibfnamefont {S.~James}\
  \bibnamefont {Allen}}, \bibinfo {author} {\bibfnamefont {Bharat}\
  \bibnamefont {Jalan}}, \bibinfo {author} {\bibfnamefont {SungBin}\
  \bibnamefont {Lee}}, \bibinfo {author} {\bibfnamefont {Daniel~G.}\
  \bibnamefont {Ouellette}}, \bibinfo {author} {\bibfnamefont {Guru}\
  \bibnamefont {Khalsa}}, \bibinfo {author} {\bibfnamefont {Jan}\ \bibnamefont
  {Jaroszynski}}, \bibinfo {author} {\bibfnamefont {Susanne}\ \bibnamefont
  {Stemmer}}, \ and\ \bibinfo {author} {\bibfnamefont {Allan~H.}\ \bibnamefont
  {MacDonald}},\ }\bibfield  {title} {\enquote {\bibinfo {title}
  {{Conduction-band edge and Shubnikov--de Haas effect in low-electron-density
  SrTiO$_3$}},}\ }\href@noop {} {\bibfield  {journal} {\bibinfo  {journal}
  {Phys. Rev. B}\ }\textbf {\bibinfo {volume} {88}},\ \bibinfo {pages} {045114}
  (\bibinfo {year} {2013})}\BibitemShut {NoStop}%
\bibitem [{SM()}]{SM}%
  \BibitemOpen
  \href@noop {} {}\bibinfo {note} {See supplemental material for a detailed
  description of model.}\BibitemShut {Stop}%
\bibitem [{\citenamefont {Hong}\ and\ \citenamefont
  {Vanderbilt}(2011)}]{Hong:2011hc}%
  \BibitemOpen
  \bibfield  {author} {\bibinfo {author} {\bibfnamefont {Jiawang}\ \bibnamefont
  {Hong}}\ and\ \bibinfo {author} {\bibfnamefont {David}\ \bibnamefont
  {Vanderbilt}},\ }\bibfield  {title} {\enquote {\bibinfo {title}
  {{First-principles theory of frozen-ion flexoelectricity}},}\ }\href@noop {}
  {\bibfield  {journal} {\bibinfo  {journal} {Phys. Rev. B}\ }\textbf {\bibinfo
  {volume} {84}},\ \bibinfo {pages} {2069} (\bibinfo {year}
  {2011})}\BibitemShut {NoStop}%
\bibitem [{\citenamefont {Hong}\ \emph {et~al.}(2010)\citenamefont {Hong},
  \citenamefont {Catalan}, \citenamefont {Scott},\ and\ \citenamefont
  {Artacho}}]{Hong:2010kx}%
  \BibitemOpen
  \bibfield  {author} {\bibinfo {author} {\bibfnamefont {JiawProc.~Nat.ang}\
  \bibnamefont {Hong}}, \bibinfo {author} {\bibfnamefont {G.}~\bibnamefont
  {Catalan}}, \bibinfo {author} {\bibfnamefont {J.~F.}\ \bibnamefont {Scott}},
  \ and\ \bibinfo {author} {\bibfnamefont {E.}~\bibnamefont {Artacho}},\
  }\bibfield  {title} {\enquote {\bibinfo {title} {{The flexoelectricity of
  barium and strontium titanates from first principles}},}\ }\href@noop {}
  {\bibfield  {journal} {\bibinfo  {journal} {J. Phys. Cond. Mat.}\ }\textbf
  {\bibinfo {volume} {22}},\ \bibinfo {pages} {112201} (\bibinfo {year}
  {2010})}\BibitemShut {NoStop}%
\bibitem [{\citenamefont {Zubko}\ \emph {et~al.}(2007)\citenamefont {Zubko},
  \citenamefont {Catalan}, \citenamefont {Buckley}, \citenamefont {Welche},\
  and\ \citenamefont {Scott}}]{Zubko:2007kh}%
  \BibitemOpen
  \bibfield  {author} {\bibinfo {author} {\bibfnamefont {P.}~\bibnamefont
  {Zubko}}, \bibinfo {author} {\bibfnamefont {G.}~\bibnamefont {Catalan}},
  \bibinfo {author} {\bibfnamefont {A.}~\bibnamefont {Buckley}}, \bibinfo
  {author} {\bibfnamefont {P.~R.~L.}\ \bibnamefont {Welche}}, \ and\ \bibinfo
  {author} {\bibfnamefont {J.~F.}\ \bibnamefont {Scott}},\ }\bibfield  {title}
  {\enquote {\bibinfo {title} {Strain-gradient-induced polarization in
  {SrTiO$_3$} single crystals},}\ }\href@noop {} {\bibfield  {journal}
  {\bibinfo  {journal} {Phys. Rev. Lett.}\ }\textbf {\bibinfo {volume} {99}},\
  \bibinfo {pages} {520} (\bibinfo {year} {2007})}\BibitemShut {NoStop}%
\bibitem [{\citenamefont {Zubko}\ \emph {et~al.}(2008)\citenamefont {Zubko},
  \citenamefont {Catalan}, \citenamefont {Buckley}, \citenamefont {Welche},\
  and\ \citenamefont {Scott}}]{Zubko:2008gh}%
  \BibitemOpen
  \bibfield  {author} {\bibinfo {author} {\bibfnamefont {P.}~\bibnamefont
  {Zubko}}, \bibinfo {author} {\bibfnamefont {G.}~\bibnamefont {Catalan}},
  \bibinfo {author} {\bibfnamefont {A.}~\bibnamefont {Buckley}}, \bibinfo
  {author} {\bibfnamefont {P.~R.~L.}\ \bibnamefont {Welche}}, \ and\ \bibinfo
  {author} {\bibfnamefont {J.~F.}\ \bibnamefont {Scott}},\ }\bibfield  {title}
  {\enquote {\bibinfo {title} {Erratum: Strain-gradient-induced polarization in
  {SrTiO$_3$} single crystals [phys. rev. lett. 99, 167601 (2007)]},}\
  }\href@noop {} {\bibfield  {journal} {\bibinfo  {journal} {Phys. Rev. Lett.}\
  }\textbf {\bibinfo {volume} {100}},\ \bibinfo {pages} {199906} (\bibinfo
  {year} {2008})}\BibitemShut {NoStop}%
\bibitem [{\citenamefont {Niranjan}\ \emph {et~al.}(2009)\citenamefont
  {Niranjan}, \citenamefont {Wang}, \citenamefont {Jaswal},\ and\ \citenamefont
  {Tsymbal}}]{Niranjan:2009ix}%
  \BibitemOpen
  \bibfield  {author} {\bibinfo {author} {\bibfnamefont {Manish~K.}\
  \bibnamefont {Niranjan}}, \bibinfo {author} {\bibfnamefont {Yong}\
  \bibnamefont {Wang}}, \bibinfo {author} {\bibfnamefont {Sitaram~S.}\
  \bibnamefont {Jaswal}}, \ and\ \bibinfo {author} {\bibfnamefont {Evgeny~Y.}\
  \bibnamefont {Tsymbal}},\ }\bibfield  {title} {\enquote {\bibinfo {title}
  {{Prediction of a Switchable Two-Dimensional Electron Gas at Ferroelectric
  Oxide Interfaces}},}\ }\href@noop {} {\bibfield  {journal} {\bibinfo
  {journal} {Phys. Rev. Lett.}\ }\textbf {\bibinfo {volume} {103}},\ \bibinfo
  {pages} {1061} (\bibinfo {year} {2009})}\BibitemShut {NoStop}%
\bibitem [{\citenamefont {Wang}\ \emph {et~al.}(2010)\citenamefont {Wang},
  \citenamefont {Niranjan}, \citenamefont {Janicka}, \citenamefont {Velev},
  \citenamefont {Zhuravlev}, \citenamefont {Jaswal},\ and\ \citenamefont
  {Tsymbal}}]{Wang:2010cp}%
  \BibitemOpen
  \bibfield  {author} {\bibinfo {author} {\bibfnamefont {Yuxuan}\ \bibnamefont
  {Wang}}, \bibinfo {author} {\bibfnamefont {M.~K.}\ \bibnamefont {Niranjan}},
  \bibinfo {author} {\bibfnamefont {K.}~\bibnamefont {Janicka}}, \bibinfo
  {author} {\bibfnamefont {J.~P.}\ \bibnamefont {Velev}}, \bibinfo {author}
  {\bibfnamefont {M.~Ye}\ \bibnamefont {Zhuravlev}}, \bibinfo {author}
  {\bibfnamefont {S.~S.}\ \bibnamefont {Jaswal}}, \ and\ \bibinfo {author}
  {\bibfnamefont {E.~Y.}\ \bibnamefont {Tsymbal}},\ }\bibfield  {title}
  {\enquote {\bibinfo {title} {{Ferroelectric dead layer driven by a polar
  interface}},}\ }\href@noop {} {\bibfield  {journal} {\bibinfo  {journal}
  {Phys. Rev. B}\ }\textbf {\bibinfo {volume} {82}},\ \bibinfo {pages} {1040}
  (\bibinfo {year} {2010})}\BibitemShut {NoStop}%
\end{thebibliography}

\begin{thebibliography}{10}
\bibitem{Khalsa}{G.\ Khalsa and A. H. MacDonald, Phys. Rev. B {\bf 86}, 125121 (2012)} 
\bibitem{Allen}{S. James Allen, {\em et al.}, Phys. Rev. B {\bf 88}, 045114 (2013).} 
\bibitem{Amany}{Amany Raslan, {\em et al.}, Phys. Rev. B {\bf 95} 054106 (2017).}
\bibitem{Dec}{J. Dec, W. Kleemann, and B. Westwanski, J.~Phys.\ Cond.\ Mat.\ {\bf 11}, L379 (1999).}
\bibitem{Uwe}{H. Uwe and T. Sakudo, Phys. Rev. B {\bf 13}, 271 (1976).} 
\bibitem{Oles}{A.\ M.\ Ol\'es, Phys. Rev. B {\bf 28}, 327 (1983).} 
\bibitem{Eyert}{V. Eyert, J. Comp. Phys. {\bf 124}, 271 (1996).}
\end{thebibliography}
\end{document}